\documentclass[12pt]{article}
\usepackage{amssymb,amsmath,epsfig}
\usepackage{graphicx}
\allowdisplaybreaks

\begin{document}
\title{\bf Viable Wormhole Structures and Energy Conditions in $f(\textmd{Q},\textmd{T})$ Theory}
\author{M. Zeeshan Gul \thanks {mzeeshangul.math@gmail.com}~, M. Sharif
\thanks{msharif.math@pu.edu.pk}~, Shajee Shahid \thanks{shajeeshaid0312@gmail.com}~ and
Faisal Javed \thanks{faisaljaved.math@gmail.com}\\
Department of Mathematics and Statistics, The University of Lahore,\\
1-KM Defence Road Lahore-54000, Pakistan.\\
Department of Physics, Zhejiang Normal University, Jinhua 321004,\\
People's Republic of China.}

\date{}
\maketitle

\begin{abstract}
This paper explores static wormhole solutions in
$f(\textmd{Q},\textmd{T})$ theory, where $\textmd{Q}$ is the
non-metricity and $\textmd{T}$ is the trace of energy-momentum
tensor. We derive the field equations that describe gravitational
phenomena in the existence of non-metricity and matter source terms.
We examine different models of this theory to determine the explicit
expressions of matter contents, which are useful for analyzing the
wormhole structures. We investigate the existence of feasible
traversable wormhole solutions for constant and variable redshift
functions. To determine whether physically viable wormhole geometry
exists, we examine the graphical interpretation of energy
constraints for different values of model parameters. It is found
that realistic traversable and stable wormhole solutions exist only
for the first model of this gravity.
\end{abstract}
\textbf{Keywords:} $f(\textmd{Q},\textmd{T})$ theory; Wormhole;
Energy conditions.\\
\textbf{PACS}: 98.62.Gq, 04.50.Kd, 83C15.

\section{Introduction}

The scientific community has been fascinated by the universe mystery
and one of the most controversial subjects has been the theoretical
idea of wormholes (\textit{WHs}). Wormholes are fictitious
tunnel-like formations that link different parts of spacetime
\cite{1}. The geometry of \textit{WHs} was first formulated by using
the Schwarzschild solution in \cite{2}. Further, Einstein and Rosen
\cite{3} suggested that spacetime might be connected via a bridge
between distinct regions, termed the Einstein-Rosen bridge. The
formation of a \textit{WH} structure requires a specific spacetime
curvature induced by a sufficient amount of matter. However, exotic
matter (which violates energy conditions (\textit{ECs})) is
necessary for the stability of \textit{WH} structures. Exotic matter
is essential for stabilizing and making possible traversal of
\textit{WHs} according to theoretical analysis. Schwarzschild
\textit{WHs} are non-traversable because of the throat abrupt
expansion and contraction \cite{4}. Morris and Thorne \cite{5}
examined the first traversable \textit{WH} solution that let matter
flow through in both directions. It is important to note that the
existence of exotic matter is necessary for traversable \textit{WHs}
to remain viable and stable. In theoretical physics, there is still
discussion and interest in the study of \textit{WHs} and their
characteristics \cite{6}-\cite{8}. The maximal amount of exotic
matter in the bridge raises concerns about the viability of the
\textit{WH} construction. Therefore, for a feasible \textit{WH}
geometry, there should be a sufficient amount of exotic matter in
the bridge. Numerous techniques have been developed to examine the
feasible \textit{WH} structures \cite{10}. Many researchers studied
the viable \textit{WH} geometry through different \textit{WH} shape
functions (\textit{WSFs}) \cite{11}. These functions enhance our
understanding of the hypothetical configurations and attributes.

Traversable \textit{WHs} are fascinating theoretical constructs that
provide a unique window into understanding the interplay between
gravity and quantum phenomena. In \textit{GTR}, traversable
\textit{WHs} require exotic matter or matter with negative energy
density to remain open and stable, which violates \textit{ECs}.
However, in the context of quantum gravity, these constraints may be
relaxed or reinterpreted. In particular, quantum field theory in
curved spacetime allows for the possibility of quantum fluctuations
generating negative energy as seen in phenomena like the Casimir
effect. These quantum fluctuations suggest that traversable
\textit{WHs} might exist without exotic matter in certain quantum
gravity frameworks. In recent years, traversable \textit{WHs} have
also gained attention through their connections to quantum
information theory. This provides a framework where \textit{WHs} can
be related to entangled quantum states, offering a potential
holographic description of quantum gravity phenomena. Furthermore,
traversable \textit{WHs} serve as theoretical laboratories for
studying non-trivial spacetime topologies and the causal structure
of spacetime in quantum gravity. They offer a promising avenue for
investigating the nature of spacetime at the Planck scale and could
lead to insights into fundamental issues such as the resolution of
singularities, the unitarity problem in black hole evaporation and
the possible existence of shortcuts through spacetime. This
highlights how traversable \textit{WHs} not only serve as
speculative solutions in \textit{GTR} but also play a significant
role in advancing our understanding of quantum gravity, particularly
in scenarios where classical constraints are reinterpreted through
quantum mechanical principles.

Traversable \textit{WHs} relate the geometry of spacetime (described
by the metric) to the matter and energy present (described by the
stress-energy tensor) have an interesting relationship with the
\textit{ECs}. For a \textit{WH} to remain open and traversable, the
geometry needs to avoid forming singularities or event horizons that
would trap travelers inside. various energy conditions are used to
describe reasonable physical matter distributions. Energy conditions
are sets of constraints on the stress-energy tensor that aim to
describe normal matter, or exotic matter. When discussing
traversable \textit{WHs}, the focus is primarily on the violation of
these \textit{ECs}. For a \textit{WH} to be traversable, the
geometry must be such that the throat of the \textit{WH} remains
open and stable. In traversable \textit{WHs}, it is generally
required that the energy conditions are violated in regions around
the throat. This implies the presence of exotic matter, allowing the
throat to remain open for passage. The energy condition violations
in traversable \textit{WHs} are often associated with exotic matter,
allowing the \textit{WH} to maintain its open structure. This exotic
matter counteracts the attractive nature of gravity, preventing the
collapse of the \textit{WH} throat. Thus, traversable \textit{WHs}
require violations of various \textit{ECs} to ensure their stability
and traversability. The presence of exotic matter plays a key role
in this violation.

Einstein general theory of relativity is serves as a cornerstone by
providing a comprehensive description of gravitational field and
matter on cosmic scales. In \textit{GTR}, spacetime is described
using mathematical structures defined by Riemann metric. This metric
encodes information about distance and angles in spacetime which
allow physicists to understand the curvature of spacetime caused by
gravity. Weyl \cite{12} introduced a more general geometric
framework as an extension of \textit{GTR}. Weyl theory incorporates
the concept of a \emph{length connection} which differs from the
standard metric connection used in Riemannian geometry
(\textit{RG}). Weyl theory focuses on gauging the conformal factor
adjusting the scale of distances. Weyl introduced the concept of
non-metricity which assures that covariant derivative of metric
tensor exists. Non-metricity and torsion are two major concepts of
non-Riemannian geometries. Alternative theories include torsion and
non-metricity as additional geometrical characteristics of spacetime
whereas Einstein formulation of \textit{GTR} emphasizes curvature.
There are two equivalent geometric representations of \textit{GTR},
i.e., the curvature representation vanishes torsion and
non-metricity, whereas the teleparallel representation vanishes
curvature and non-metricity. However, one more comparable
representation of the geometric features is the non-metricity of the
metric.

In the framework of \textit{GTR}, the gravitational interaction is
described by a purely metric theory where the connection is assumed
to be Levi-Civita, implying zero non-metricity. However, in
alternative theories of gravity, especially those attempting to
unify gravity with other forces or explain phenomena such as dark
energy, dark matter, or early universe cosmology, the assumption of
a purely metric theory is relaxed. Non-metricity arises in such
scenarios and can be an essential feature of extended geometrical
frameworks such as the metric-affine gravity or Teleparallel
Gravity. Non-metricity describes how the length of vectors changes
under parallel transport and it is closely linked to the concept of
varying gravitational couplings or connections that include degrees
of freedom beyond the metric. These modifications can provide a deep
understanding of phenomena that \textit{GTR} struggles to explain
such as the nature of singularities, the behavior of gravity at
quantum scales and potential resolutions to cosmological problems
like the singularity in the Big Bang or cosmological bounce models.
Thus, studying non-metricity offers a promising pathway to explore
gravitational theories that extend beyond the well-established
predictions of \textit{GTR}, opening doors to resolving some of the
fundamental issues in modern cosmology and gravitational physics.

Studying non-metricity is motivated by several key factors as
\begin{itemize}
\item
Extensions to the Geometric Nature of Gravity
\end{itemize}
Non-metricity introduces an additional layer of geometry that
departs from the constraints of \textit{GTR}. While \textit{GTR} is
based on the curvature of spacetime, incorporating non-metricity
allows for a more generalized theory of spacetime that includes
non-metricity as fundamental geometric entity. This geometric
extension provides alternative mechanisms for understanding
gravitational interactions.
\begin{itemize}
\item
Unification with Quantum Gravity
\end{itemize}
A persistent challenge in theoretical physics is reconciling
\textit{GTR} with quantum mechanics. Non-metricity may offer a
pathway toward quantum gravity by introducing new symmetries or
degrees of freedom that bridge the gap between classical gravity and
quantum field theory. By modifying the fundamental structure of
spacetime, theories with non-metricity have the potential to
accommodate quantum effects that \textit{GTR} cannot explain such as
those predicted by string theory or loop quantum gravity.
\begin{itemize}
\item
Exploring Dark Energy and Dark Matter
\end{itemize}
The introduction of non-metricity may also provide insight into the
nature of dark energy and dark matter, which remain elusive in the
context of \textit{GTR}. Non-metricity could introduce additional
fields or modifications to spacetime dynamics that mimic the effects
of these unknown components of the universe. It also offers a
possible explanation for the accelerated expansion of the universe
and the missing mass inferred from galactic rotation curves without
needing to invoke unknown particles.
\begin{itemize}
\item
High-Energy Regimes and Early Universe Cosmology
\end{itemize}
In the early universe, conditions of extremely high energy and
curvature likely deviate from the predictions of \textit{GTR}.
Theories that incorporate non-metricity such as teleparallel gravity
or generalized affine theories could provide a more accurate
description of the early universe, resolving issues like the big
bang singularity and offering alternatives to cosmic inflation.
\begin{itemize}
\item
The Role in Modified Gravity Theories
\end{itemize}
Non-metricity is a key feature in several modified gravity theories,
including the metric-affine gravity framework and theories of
gravity that generalize the Palatini formalism. These approaches
offer solutions to long-standing issues in cosmology and
astrophysics such as the cosmological constant problem and the need
for new forms of energy to explain gravitational phenomena at large
scales. Hence, the study of non-metricity extends the geometric
framework of gravity, providing a fertile ground for new insights
into gravitational phenomena that \textit{GTR} cannot fully address.
By incorporating non-metricity, researchers aim to uncover a more
complete theory of gravity that not only explains known phenomena
but also sheds light on the mysteries of dark matter, dark energy,
quantum gravity and the early universe.

Teleparallel theory is one such alternative theory where torsion
represents the gravitational interaction. In symmetric teleparallel
theory, the gravitational interaction is represented by
non-metricity. To characterize \textit{GTR} in the context of
torsion and non-metricity, the integral actions are expressed as
$\int\sqrt{-g}\mathcal{T}$ \cite{13a} and $\int\sqrt{-g}\textmd{Q}$
\cite{13b} where $\mathcal{T}$ represents torsion and $\textmd{Q}$
represents non-metricity, respectively. Yixin et al \cite{13c}
formulated the $f(\textmd{Q},\textmd{T})$ theory by assimilating the
trace of stress-energy tensor in the functional action of
$f(\textmd{Q})$ gravity. The motivation behind this theory is to
examine theoretical impacts by observational data and cosmological
domains. Xu et al \cite{13e} developed the Weyl form of
$f(\textmd{Q},\textmd{T})$ gravity and its implications in the
background of cosmology. Arora et al \cite{13f} investigated the
cosmic acceleration in the absence of dark energy. Bhattacharjee
\cite{13ff} explored the applications of $f(\textmd{Q},\textmd{T})$
gravity and found that this gravity alter the nature of tidal forces
and equation of motion in Newtonian limit. Pati et al \cite{13g}
established a mathematical framework in this theory in terms of the
Hubble model. Agrawal et al \cite{13h} presented the dynamical
features and matter bounce scenario in this background. Shiravand et
al \cite{13i} explained the cosmic inflation in the same theory.

The $f(\textmd{Q},\textmd{T})$ theory of gravity is a modified
theory that generalizes the symmetric teleparallel gravity, focusing
on the non-metricity scalar and the trace of the energy-momentum
tensor. In this framework, non-metricity represents a geometrical
object that quantifies how the length of vectors changes during
parallel transport. The trace of the energy-momentum tensor includes
the contribution from matter fields. This theory departs from the
\textit{GTR} and other modified theories by introducing coupling
between geometry (non-metricity) and matter field (trace of the
energy-momentum tensor), explaining more accurately cosmological and
astrophysical phenomena.
\begin{itemize}
\item
Comparison with Other Modified Gravity Theories
\end{itemize}
Einstein general theory of relativity is based on the curvature of
spacetime with the Ricci scalar as the fundamental geometrical
object. The $f(\textmd{R})$ gravity is one of the most well-known
modifications to \textit{GTR}, where the Ricci scalar in the
Einstein-Hilbert action is replaced by a function $f(\textmd{R})$,
allowing more complex interactions between curvature and gravity.
The field equations are fourth-order differential equations instead
of the second-order equations in \textit{GTR}. The
$f(\textmd{R},\textmd{T})$ gravity extends $f(\textmd{R})$ theory by
introducing a dependence on both the Ricci scalar and the trace of
the energy-momentum tensor, allowing a coupling between geometry and
matter. The equations of motion are more complex than
$f(\textmd{R})$ gravity and lead to non-conservation of the
energy-momentum tensor. Teleparallel gravity reformulates
\textit{GTR} by using torsion rather than curvature to describe
gravity. The gravitational action is constructed from the torsion
scalar. In teleparallel gravity, the field equations involve the
torsion tensor rather than the curvature of spacetime. Scalar-tensor
theories introduce a scalar field with the tensor field (the metric)
and gravity is described by both. The equations of motion involve
both the metric and the scalar field, leading to a richer set of
dynamics, including varying gravitational coupling. The higher-order
theories extend \textit{GTR} by adding terms involving higher powers
of curvature such as the Gauss-Bonnet term, which is a specific
combination of curvature invariants. The corresponding field
equations lead to more complex field equations but retaining
second-order derivatives in the Gauss-Bonnet case.

In $f(\textmd{Q},\textmd{T})$, the gravitational interaction is
based on non-metricity rather than curvature with additional terms
involving the matter trace, which introduces modifications to how
gravity couples to matter. Both $f(\textmd{Q},\textmd{T})$ and
$f(\textmd{R})$ theories generalize \textit{GTR}, but
$f(\textmd{Q},\textmd{T})$ is based on non-metricity rather than
curvature. Additionally, $f(\textmd{Q},\textmd{T})$ includes the
matter-energy trace, creating a more direct interaction between
matter and geometry. While $f(R,\textmd{T})$ theory involves
coupling between geometry and matter via the trace of
energy-momentum tensor, the geometrical foundation differs is based
on non-metricity in f(\textmd{Q},\textmd{T}) theory. This gives a
distinct structure and different implications for cosmology and
gravity. The $f(\textmd{Q},\textmd{T})$ does not involve an
additional scalar field but modifies gravity directly through the
non-metricity scalar. The coupling with $\textmd{T}$ in
$f(\textmd{Q},\textmd{T})$ introduces effective modifications
similar to scalar fields but without introducing new fundamental
fields. The $f(\textmd{Q},\textmd{T})$ theory can explain the
late-time acceleration of the universe without invoking a
cosmological constant or dark energy. The coupling between
$\textmd{Q}$ and $\textmd{T}$ offers a novel approach to modifying
gravity, allowing for an effective cosmological constant that varies
with the matter content of the universe.

The study of \textit{WH} geometry in modified theories has become
subject of great interest for cosmologists in the recent years. Lobo
et al \cite{14} investigated traversable \textit{WH} structures
using various WSFs in $f(\textmd{R})$ theory. Mazharimousavi and
Halilsoy \cite{15} discovered that \textit{WH} solutions meet all
required viability conditions near the \textit{WH} throat in this
theory. The traversable \textit{WH} geometry through Noether
symmetry in the context of scalar-tensor theory has been studied in
\cite{16}. The static \textit{WH} solutions through \textit{ECs} in
$f(\textmd{R},\textmd{T})$ has been examined in \cite{17}. In
$f(\textmd{R})$ theory, the viable \textit{WH} solutions using
Noether symmetry have been studied in \cite{18}. Mustafa et al
\cite{19} used the Karmarkar condition to examine the viable
\textit{WH} geometry in $f(\textmd{Q})$ theory. Using the embedding
class-I technique in $f(\textmd{R})$ theory, Shamir and Fayyaz
\cite{20} created a \textit{WSF} and found \textit{WH} structure
with a small amount of exotic matter. It was discovered that
\textit{WH} solutions which match the linear and exponential models
of $f(\textmd{Q})$ gravity models are stable and physically feasible
\cite{21}. The Karmarkar condition was applied in \cite{22} to
analyze traversable \textit{WH} structure in $f(\textmd{R})$ theory.
Gul and his collaborators \cite{22a} developed \textit{WSF} through
Karmarkar constraint to examine the geometry of \textit{WH}
structures in different modified theories of gravity. Recently, the
study of observational constraints in modified gravities discussed
in \cite{23}-\cite{25}.

This paper investigates viable traversable \textit{WH} solutions
using the embedding class-I technique in $f(\textmd{R},\textmd{G})$
theory. The analysis focuses on studying the behavior of shape
function and \textit{ECs} in this context. Wormholes are intriguing
solutions to the Einstein field equations that have captured
significant attention due to their implications in cosmology and
interstellar travel. However, their viability and stability in the
framework of alternative gravitational theories remain an open
question. The motivation for exploring $f(\textmd{R},\textmd{G})$
gravity is twofold. First, this gravitational theory is an extension
of \textit{GTR} that allows for a more comprehensive description of
gravitational phenomena. Second, \textit{WH} solutions in
$f(\textmd{R},\textmd{G})$ gravity offer new insights between
gravity modifications and exotic structures like \textit{WHs}. By
investigating \textit{WHs} in this modified gravity theory, we aim
to contribute our understanding on the existence and stability of
\textit{WH} solutions. Furthermore, investigating \textit{WHs} in
this framework may shed light on the compatibility of \textit{WHs}
with modified gravity theories, which has implications for
theoretical physics and observational cosmology.

To our knowledge, there has been limited exploration of \textit{WHs}
in the context of $f(\textmd{Q},\textmd{T})$ gravity. Our study
takes a pioneering step in examining the existence and stability of
\textit{WH} solutions in this specific modified theory. Most
previous research on \textit{WHs} in modified theories focused on
$f(\textmd{Q})$ gravity. In contrast, our approach considers the
joint effects of the non-metricity scalar and the trace of the
stress-energy tensor in $f(\textmd{Q},\textmd{T})$ gravity,
providing a more comprehensive analysis. We intend to perform a
detailed viability analysis of the \textit{WH} solutions in
$f(\textmd{Q},\textmd{T})$ gravity, which is a novel aspect of our
study. By conducting a comprehensive analysis of \textit{WH}
solutions in $f(\textmd{Q},\textmd{T})$ gravity, our work
contributes to the broader understanding of gravitational theories
and their astrophysical implications.

The literature mentioned above encourages us to investigate
\textit{WH} geometry in $f(\textmd{Q},\textmd{T})$ theory.  This
paper follows the following pattern. In section \textbf{2}, we
provide the fundamental formulation of the field equations of
$f(\textmd{Q},\textmd{T})$ theory. The field equations corresponding
to Morris-Thorne spacetime in $f(\textmd{Q},\textmd{T})$ theory are
developed in section \textbf{3}. The feasible \textit{WH} geometry
corresponding to several $f(\textmd{Q},\textmd{T})$ models with
various shape functions and a constant redshift function is
investigated in section \textbf{4}. Our results are summarized in
the final section.

\section{$f(\textmd{Q},\textmd{T})$ Theory-Basic Formalism}

This section describes the basics of modified
$f(\textmd{Q},\textmd{T})$ gravity and formulates the field
equations using the variational method. Weyl \cite{12} generalized
\textit{RG} based on the assumption that an arbitrary vector
undergoes a change in length during parallel transport. Accordingly,
the fundamental fields of Weyl space is represented by a new vector
field $(w^{\alpha})$ and the metric tensor. The expression $\delta
l=lw_{\alpha} \delta x^{\alpha}$ \cite{32} represents the change in
length of a vector transported along an infinitesimal path in Weyl
space. Additionally, the explanation for the variance in the vector
length in Weyl space resulting from parallel transport is
\begin{equation}\label{1}
\delta l = l(\nabla_{\beta}\omega_{\alpha}-\nabla_{\alpha}
\omega_{\beta})s^{\alpha\beta}.
\end{equation}
If there is a local scaling length given by $\tilde{l} = \sigma(x)l$
then it changes the vector field $w_{\alpha}$ to
$\tilde{w_{\alpha}}=w_{\alpha}+(ln\sigma),_{\alpha}$. Moreover, the
elements of the metric tensor transform under conformal
transformations as
$\tilde{g}{\alpha\beta}=\sigma^{2}{g}_{\alpha\beta}$ and
$\tilde{g}^{\alpha\beta}=\sigma^{-2}{g}^{\alpha\beta}$ \cite{33}.
The Weyl geometry also contains a semi-metric connection, defined as
\begin{equation}\label{2}
\bar{\Gamma}^{\theta}_{\alpha\beta}=\Gamma^{\theta}_{~\alpha\beta}
+g_{\alpha\beta}\omega^{\theta}-\delta^{\theta}_{\alpha}\omega_{\beta}
-\delta^{\theta}_{\beta}\omega_{\alpha},
\end{equation}
where $\Gamma^{\theta}_{\alpha\beta}$ represents the Christoffel
symbol. The gauge covariant derivative can be formulated using the
fact that $\bar{\Gamma}^{\theta}_{\alpha\beta}$ is symmetric. The
Weyl tensor is expressed as
\begin{equation}\label{3}
\bar{\textmd{R}}_{\alpha\beta\gamma\eta}=\bar{\textmd{R}}
_{(\alpha\beta)\gamma\eta}
+\bar{\textmd{R}}_{[\alpha\beta]\gamma\eta},
\end{equation}
where
\begin{eqnarray}\nonumber
\bar{\textmd{R}}_{[\alpha\beta]\gamma\eta}&=&\textmd{R}
_{\alpha\beta\gamma\eta}
+2\nabla_{\gamma}~\omega_{[\alpha}g_{\beta]}\eta
+2\nabla_{\eta}~\omega_{[\beta}g_{\alpha]}\gamma +2\omega_{\gamma}
\omega_{[\alpha}g_{\beta]}\eta
\\\label{4}
&+&2\omega_{\eta}\omega_{[\beta}g_{\alpha]}\gamma
-2\omega^{2}g_{\gamma}
{[_{\alpha}}g_{\beta}]\eta,
\end{eqnarray}
and
\begin{equation}\label{5}
\bar{\textmd{R}}_{(\alpha\beta)\gamma\eta}
=\frac{1}{2}(\bar{\textmd{R}} _{\alpha\beta\gamma\eta}
+\bar{\textmd{R}}_{\beta\alpha\gamma\eta})
=g_{\alpha\beta}W_{\gamma\eta}.
\end{equation}
The first contraction of the Weyl curvature tensor yields
\begin{equation}\label{6}
\bar{\textmd{R}}^{\alpha}_{~\beta}=\bar{\textmd{R}}
^{\gamma\alpha}_{~\gamma\beta}=
{\textmd{R}}^{\alpha}_{~\beta}+2\omega^{\alpha}\omega
_{\beta}+3\nabla_{\beta}\omega^{\alpha}
-\nabla_{\alpha}\omega^{\beta}+g^{\alpha}_{~\beta}
(\nabla_{\gamma}\omega^{\gamma}-2\omega_{\gamma} \omega^{\gamma}).
\end{equation}
Finally, the Weyl scalar is obtained as
\begin{equation}\label{7}
\bar{\textmd{R}}=\bar{\textmd{R}}^{\gamma}_{~\gamma}
=\textmd{R}+6(\nabla_{\alpha}\omega^{\alpha}-\omega
_{\alpha}\omega^{\alpha}).
\end{equation}

Compared to \textit{RG} and Weyl geometry, Weyl-Cartan spaces with
torsion provided a more generalized geometric framework. The term
torsion describes the antisymmetric portion of the connection, which
is a measurement of the connection twisting related to parallel
transport. The law of parallel transport is given by
$d\upsilon^{\alpha}=-\upsilon^{\sigma}\hat\Gamma^{\alpha}
_{\sigma\beta}dx^{\beta}$ \cite{34}. In Weyl-Cartan geometry, the
connection is defined as
\begin{equation}\label{8}
\hat{\Gamma}^{\theta}_{~\alpha\beta}=\Gamma^{\theta}_{\alpha\beta}
+\textmd{C}^{\theta}_{~\alpha\beta}+L^{\theta}_{~\alpha\beta},
\end{equation}
where
\begin{eqnarray}\label{10}
\textmd{C}^{\theta}_{~\alpha\beta}=\hat{\Gamma}^{\theta}
_{~[\alpha\beta]}+g^{\theta\sigma}
g_{\alpha\kappa}\hat{\Gamma}^{\kappa}_{~[\beta\sigma]}
+g^{\theta\sigma}g_{\beta\kappa}\hat{\Gamma}^{\kappa}
_{~[\alpha\sigma]},
\end{eqnarray}
is the contortion tensor and the deformation tensor is expressed as
\begin{equation}\label{11}
L^{\theta}_{\alpha\beta}=\frac{1}{2}g^{\theta\sigma}
(\textmd{Q}_{\alpha\beta\sigma}
+\textmd{Q}_{\beta\alpha\sigma}-\textmd{Q}_{\alpha\sigma\beta}),
\end{equation}
with
\begin{equation}\label{12}
\textmd{Q}_{\theta\alpha\beta}=-\frac{\partial{g}_{\alpha\beta}}
{\partial\chi^{\theta}}+g_{\beta\sigma}\hat{\Gamma}
^{\sigma}_{\alpha\theta}+g_{\sigma\alpha}
\hat{\Gamma}^{\sigma}_{~\beta\theta}.
\end{equation}
Using Eq.(\ref{8}) with
$\textmd{Q}_{\theta\alpha\beta}=-2g_{\alpha\beta}\omega_{\theta}$,
we have
\begin{equation}\label{13}
\hat{\Gamma}^{\theta}_{\alpha\beta}={\Gamma}^{\theta}
_{~\alpha\beta}+g_{\alpha\beta}
\omega^{\theta}-\delta^{\theta}_{~\alpha}\omega_{\beta}
-\delta^{\theta}_{~\beta}
\omega_{\alpha}+C^{\theta}_{~\alpha\beta},
\end{equation}
where
\begin{equation}\label{14}
C^{\theta}_{~\alpha\beta}=T^{\theta}_{~\alpha\beta}
-g^{\theta\eta}g_{\sigma\alpha}
T^{\sigma}_{~\eta\beta}-g^{\theta\eta}g_{\sigma\beta}
T^{\sigma}_{~\eta\alpha}.
\end{equation}
The Weyl-Cartan tensor is given by
\begin{equation}\label{15}
\hat{\textmd{R}}^{\theta}_{\alpha\beta\sigma}
=\hat{\Gamma}^{\theta}_{~\alpha\sigma},
_{\beta}-\hat{\Gamma}^{\theta}_{~\alpha\beta},
_{\sigma}+\hat{\Gamma}^{\gamma}_{~\alpha\sigma}
\hat{\Gamma}^{\theta}_{\gamma\beta}-\hat{\Gamma}
^{\gamma}_{\alpha\beta} \hat{\Gamma}^{\theta}_{\gamma\sigma},
\end{equation}
and
\begin{eqnarray}\nonumber
\hat{R}&=&\hat{R}^{\alpha\beta}_{~~\alpha\beta}
=R+6\nabla_{\beta}\omega^{\beta}
-4\nabla_{\beta}T^{\beta}-6\omega_{\beta}\omega
^{\beta}+8\omega_{\beta}T^{\beta}
\\\label{16}
&+&T^{\alpha\gamma\beta}T_{\alpha\gamma\beta}
+2T^{\alpha\gamma\beta}T_{\beta\gamma\alpha} -4T_{\beta}T^{\beta}.
\end{eqnarray}
By eliminating the boundary terms in the Ricci scalar, we can
reformulate the gravitational action with coupling constant one as
\cite{34a}
\begin{equation}\label{18}
S=\frac{1}{2}\int g^{\alpha\beta}(\Gamma^{\gamma}_{\sigma\alpha}
\Gamma^{\sigma}_{\beta\gamma}
-\Gamma^{\gamma}_{\sigma\gamma}\Gamma^{\sigma}
_{\alpha\beta})\sqrt{-g}d^{4}x.
\end{equation}
Using the assumption that the connection is symmetric, the
gravitational action becomes
\begin{equation}\label{19}
S=-\frac{1}{2}\int
g^{\alpha\beta}(L^{\gamma}_{\sigma\alpha}L^{\sigma}
_{\beta\gamma}-L^{\gamma}_{~\sigma\gamma}
\Gamma^{\sigma}_{~\alpha\beta})\sqrt{-g}d^{4}x.
\end{equation}
This is the action of symmetric teleparallel gravity. However, these
two gravitational theories differ in important ways. Because of the
curvature tensor disappearance, the total spacetime geometry in the
context of symmetric teleparallel gravity is flat. This leads to the
adoption of the Weitzenbock configuration in the global geometry.
Moreover, changes in the length of the vector during parallel
transit cause gravitational effects rather than the rotation of the
angle between the two vectors.

Now, we consider an extension of Eq.(\ref{19}) as
\begin{equation}\label{20}
S=\int\big(\frac{1}{2}f(\textmd{Q},\textmd{T})+\textmd{L}_{M}\big)
\sqrt{-g}d^{4}x,
\end{equation}
where
\begin{equation}\label{21}
\textmd{Q}\equiv-g^{\alpha\beta}(L^{\gamma}_{~~\eta\alpha}
L^{\eta}_{~\beta\gamma}-L^{\gamma}
_{~\eta\gamma}L^{\eta}_{~~\alpha\beta}),
\end{equation}
and
\begin{equation}\label{22}
L^{\gamma}_{~~\eta\lambda}=-\frac{1}{2}g^{\gamma\theta}
(\nabla_{\lambda}g_{\eta\theta}+\nabla_{\eta}
g_{\theta\lambda}-\nabla_{\theta} g_{\eta\lambda}).
\end{equation}
The superpotential is defined as
\begin{eqnarray}\label{24}
P^{\gamma}_{~\alpha\beta}&\equiv&\frac{1}{4}[-\textmd{Q}^{\gamma}
_{~~\alpha\beta}+2\textmd{Q}^{\quad\gamma}_{(\alpha\quad \beta)}
+\textmd{Q}^{\gamma}g_{\alpha\beta}-\hat{\textmd{Q}}^{\gamma}g_{\alpha\beta}
-\delta^{\gamma}_{~(\alpha} \textmd{Q}_{\beta)}]
\\\nonumber
&=&-\frac{1}{2}L^{\gamma}_{~~\alpha\beta}+\frac{1}{4}
(\textmd{Q}^{\gamma}-\tilde{\textmd{Q}}^{\gamma})g_{\alpha\beta}
-\frac{1}{4}\delta^{\gamma}_{~~(\alpha}\textmd{Q}_{\beta)}].
\end{eqnarray}
We obtain the relation for non metricity as
\begin{equation}\label{25}
\textmd{Q}=-\textmd{Q}_{\sigma\gamma\beta}P^{\sigma\gamma\beta}
=-\frac{1}{4}(-\textmd{Q}^{\gamma\beta\rho}
\textmd{Q}_{\gamma\beta\rho}+2\textmd{Q}^{\gamma\beta\rho}
\textmd{Q}_{\rho\gamma\beta}
-2\textmd{Q}^{\rho}\tilde{\textmd{Q}}_{\rho}+{\textmd{Q}}^{\rho}{\textmd{Q}}_{\rho}).
\end{equation}
The calculation of this relation is given in \cite{13c}. The
variation of Eq.(\ref{20}) corresponding to the metric tensor yields
\begin{eqnarray}\nonumber
\delta S&=&\int\frac{1}{2} \delta
[f(\textmd{Q},\textmd{T})\sqrt{-g}]
\delta[\textmd{L}_{M}\sqrt-g]d^{4}x,
\\\nonumber
&=&\int\frac{1}{2}\big(-\frac{1}{2}fg_{\alpha\beta}\sqrt-g\delta
g^{\alpha\beta}+f_{\textmd{Q}}\sqrt{-g}\delta
\textmd{Q}+f_{\textmd{T}}\sqrt-g\delta \textmd{T}\big)
\\\label{26}
&-&\frac{1}{2}\textmd{T}_{\alpha\beta}\sqrt-g\delta
g^{\alpha\beta}d^{4}x.
\end{eqnarray}
The detailed variation of non metricity is provided in \cite{13c}.
Furthermore, we define
\begin{equation}\label{27}
\textmd{T}_{\alpha\beta}=-\frac{2}{\sqrt-g}\frac{(\delta\sqrt{-g}
{\textmd{L}}_{M})}{\delta g^{\alpha\nu}} ,\quad
\Theta_{\alpha\beta}\equiv g^{\gamma\eta}\frac{\delta
\textmd{T}_{\gamma\eta}}{\delta g^{\alpha\beta}},
\end{equation}
which implies that $\delta
\textmd{T}=\delta(\textmd{T}_{\alpha\beta}g^{\alpha\beta})=(\textmd{T}_{\alpha\beta}
+\Theta_{\alpha\beta})\delta g^{\alpha\beta}$. Thus, Eq.(\ref{27})
becomes
\begin{eqnarray}\nonumber
\delta S&=&\int\frac{1}{2}\big[-\frac{1}{2}fg_{\alpha\beta}
\sqrt{-g}\delta
g^{\alpha\beta}+f_{\textmd{T}}(\textmd{T}_{\alpha\beta}+\Theta_{\alpha\beta})
\sqrt{-g}\delta g^{\alpha\beta}\
\\\nonumber
&-&f_{\textmd{Q}}\sqrt{-g}(P_{\alpha\gamma\eta}\textmd{Q}_{\beta}
^{~~\gamma\eta}
-2\textmd{Q}^{\gamma\eta}_{~~\alpha}P_{\gamma\eta\beta})\delta
g^{\alpha\beta}+2f_{\textmd{Q}}\sqrt{-g}P_{\gamma\alpha\beta}
\nabla^{\gamma}\delta g^{\alpha\beta}\big]
\\\label{28}
&-&\frac{1}{2}\textmd{T}_{\alpha\beta}\sqrt{-g}\delta
g^{\alpha\beta}d^{4}x.
\end{eqnarray}
The resulting modified field equations are
\begin{eqnarray}\nonumber
\textmd{T}_{\alpha\beta}&=&-\frac{2}{\sqrt-g}\nabla_{\gamma}(f_{\textmd{Q}}\sqrt-g
P^{\gamma}_{\alpha\beta})-\frac{1}{2}fg_{\alpha\beta}
+f_{\textmd{T}}(\textmd{T}_{\alpha\beta}+\Theta_{\alpha\beta})
\\\label{29}
&-&f_{\textmd{Q}}(P_{\alpha\gamma\eta}\textmd{Q}_{\beta}^{~~\gamma\eta}
-2\textmd{Q}^{\gamma\eta}_{~~\alpha}P_{\gamma\eta\beta}),
\end{eqnarray}
where $f_{\textmd{T}}=\frac{\partial f}{\partial \textmd{T}}$ and
$f_{\textmd{Q}}=\frac{\partial f}{\partial \textmd{Q}}$. This
modified framework provides insights into the behavior of gravity
through the solution of these field equations.

\section{Wormhole and Energy Conditions}

We consider the Morris-Thorne spacetime as \cite{5}
\begin{equation}\label{30}
ds^{2}=dt^{2}e^{2\mu(r)}-dr^{2}\left(1-\frac{\nu(r)}{r}\right)^{-1}
-d\theta^{2}r^{2}-d\phi^{2}r^{2}\sin^{2}\theta,
\end{equation}
where $\mu(r)$ defines the redshift function and $\nu(r)$ represents
the shape function. The given constraints must be fulfilled for a
viable \textit{WH} geometry.
\begin{itemize}
\item
$\nu(r)< r$,
\item
$\nu(r_{0})-r=0$ at $r_{0}$,
\item
$\nu'(r_{0})<1$,
\item
$\frac{\nu(r)}{r}\rightarrow0$ as $r\rightarrow\infty$,
\end{itemize}
where $r_{0}$ is the radius of $\mathrm{\textit{WH}}$ throat. We
assume fluid distribution as
\begin{equation}\label{31}
\textmd{T}_{\alpha\beta}=(\rho+p_{\bot})u_{\alpha}u_{\beta}-p_{\bot}g_{\alpha\beta}
+(p_{r}-p_{\bot})v_{\alpha\beta},
\end{equation}
where the four-velocity and four-vector are represented by the
$u_{\alpha}$ and $v_{\alpha}$, respectively. The matter-Lagrangian
density is a fundamental concept in gravitational physics that
describes the behavior and distribution of matter in a spacetime.
When the matter distribution displays distinct characteristics along
several spatial directions (anisotropic matter configuration), the
important information can be gained by examining the particular
matter-Lagrangian density. For anisotropic matter distribution, the
well-known matter-Lagrangian density in the literature is considered
as $\textmd{L}_{M}=-P=-\frac{p_{r}+2p_{\bot}}{3}$
\cite{35}-\cite{37}. The expression of $\Theta_{\alpha\beta}$ can be
expressed as
\begin{equation}\label{32}
\Theta_{\alpha\beta}=-g_{\alpha\beta}P-2\textmd{T}_{\alpha\beta}.
\end{equation}
The non-metricity scalar corresponding to Moris-Throne spacetime
turns out to be
\begin{equation}\label{33}
\textmd{Q}=-\frac{\nu}{r^2}\bigg[\frac{r\nu'-\nu}{r(r-\nu)}+2\mu'\bigg].
\end{equation}
Now, using Eqs.(\ref{29})-(\ref{33}), the resulting field equations
become
\begin{eqnarray}\nonumber
\rho&=&\frac{1}{2r^3}\bigg[f_{\textmd{Q}}\big((2r-\nu)(r
\nu'-\nu)(r-\nu)^{-1} +\nu(2r\mu'+2)\big)+2\nu
rf_{\textmd{Q}\textmd{Q}}f_{\textmd{Q}}
\\\label{36}
&+&fr^{3}-2r^{3}f_{\textmd{T}}(P+\rho)\bigg],
\\\nonumber
p_{r}&=&\frac{-1}{2r^3}\bigg[f_{\textmd{Q}}\big(\nu((r
\nu'-\nu)(r-\nu)^{-1} +2r\mu'+2)-4r\mu'\big)+2\nu
rf_{\textmd{Q}\textmd{Q}}f_{\textmd{Q}}
\\\label{37}
&+&fr^{3}+2r^{3}f_{\textmd{T}}(P-p_{r})\bigg],
\\\nonumber
p_{\bot}&=&\frac{-1}{4r^{2}}\bigg[f_{\textmd{Q}}\big\{\big((r
\nu'-\nu)(\frac{2r}{r-\nu}+2r\mu')\big)(r)^{-1}+4(2\nu-r)\mu'
-4r\mu'^{2}
\\\label{38}
&-&4r\mu''\big\}-4rf_{\textmd{Q}\textmd{Q}}f_{\textmd{Q}}\mu'+2fr^{2}{r-\nu}
-4r^{2}f_{\textmd{T}}(P-p_{\bot})\bigg].
\end{eqnarray}
The field equations are complicated because multivariate functions
and their derivatives are present. We take a specific model with
$f(\textmd{T})=b\textmd{T}$ as
\begin{equation}\label{38a}
f(\textmd{Q},\textmd{T})=f(\textmd{Q})+b\textmd{T},
\end{equation}
to simplify the field equations and obtain explicit expressions for
the pressure components and energy density. Here, $b$ is an
arbitrary constant. The field equations (\ref{36})-(\ref{38})
corresponding to this model become
\begin{eqnarray}\nonumber
\rho
&=&\big[r^{2}\big\{5br\mu'(2r\textmd{Q}'f_{\textmd{Q}\textmd{Q}}-f_{\textmd{Q}}(\nu'-4))
+r^{2}(10b\mu''f_{\textmd{Q}}-3(b+1)f_{\textmd{Q}})
\\\nonumber
&+&10br^{2}\mu'^{2}f_{\textmd{Q}}+2f_{\textmd{Q}}(2 b-3)\nu'\big\}+r
\nu\big\{-f_{\textmd{Q}}\big(20br^{2}\mu''+20br^{2}\mu'^{2}
\\\nonumber
&+&r\mu'(-5b\nu'+41b+6)+(7b-3)\nu'\big)-2rQ'(10br\mu'-2b+3)f_{\textmd{Q}\textmd{Q}}
\\\nonumber
&+&3(b+1)r^{2}f_{\textmd{Q}}\big\}+\nu^{2}\big\{r\big(10br\mu''f_{\textmd{Q}}+\mu'(10br\mu'+21b+6)f_{\textmd{Q}}
+2\textmd{Q}'
\\\nonumber
&\times&(5br\mu'-2b+3)f_{\textmd{Q}\textmd{Q}}\big)+3(b+1)f_{\textmd{Q}}\big\}\big]
\big[6r^{3}(b+1)(2b-1)
\\\label{39}
&\times& (r-\nu)\big]^{-1},
\\\nonumber
p_{r}&=&\big[r^{2}\big\{r\mu'\big(5b(f_{\textmd{Q}}\nu'-2r\textmd{Q}'f_{\textmd{Q}\textmd{Q}})+4f_{\textmd{Q}}(b-3)\big)
+r^{2}(3(b+1)f_{\textmd{Q}}
\\\nonumber
&-&10b\mu''f_{\textmd{Q}})-10br^{2}f_{\textmd{Q}}\mu'^{2}+8b\nu'f_{\textmd{Q}}\big\}+r\nu\big\{f_{\textmd{Q}}\big(20br^{2}\mu
''+20br^{2}\mu'^{2}
\\\nonumber
&+&r\mu'(-5b\nu'-7b+30)+(3-5b)\nu'\big)+(6-12b)f_{\textmd{Q}}+2r\textmd{Q}'f_{\textmd{Q}\textmd{Q}}
\\\nonumber
&\times&(10br\mu'-2b+3)-3(b+1)fr^{2}\big\}-\nu^{2}\big\{f_{\textmd{Q}}\big(r(10br\mu''+\mu'
(10cr\mu'
\\\nonumber
&-&3b+18))-9b+9\big)+2rQ'f_{\textmd{Q}\textmd{Q}}(5cr\mu'-2b+3)\big\}\big]\big[6r^{3}(b+1)
\\\label{40}
&\times&(2b-1)(r-\nu)\big]^{-1},
\\\nonumber
p_{\bot}&=&\big[r^{2}\big\{r\mu'\big(2f_{\textmd{Q}\textmd{Q}}(b-3)r
\textmd{Q}'-f_{\textmd{Q}}((b-3)\nu'+8b
+6)\big)+r^{2}(2f_{\textmd{Q}}(b-3)\mu''
\\\nonumber
&+&3(b+1)f)+2f_{\textmd{Q}}(b-3)r^{2}\mu'^{2}+f_{\textmd{Q}}(2b+3)\nu'\big\}+r\nu
\big\{f_{\textmd{Q}}\big(-4(b-3)
\\\nonumber
&\times&r^{2}\mu''-4(b-3)r^{2}\mu '^{2}+r\mu'((b-3)
\nu'+23b+15)+b\nu'\big)+f_{\textmd{Q}}(6b-3)
\\\nonumber
&+&4f_{\textmd{Q}\textmd{Q}}
r\textmd{Q}'\big(2b-(b-3)r\mu'\big)-3(b+1)
nr^{2}\big\}+\nu^{2}\big(r\big(2f_{\textmd{Q}}(b-3)r\mu''
\\\nonumber
&+&f_{\textmd{Q}} \mu'\big(2(b-3)r\mu'-3(5
b+3)\big)+2f_{\textmd{Q}\textmd{Q}}\textmd{Q}'\big((b-3)
r\mu'-4b\big)\big)-9f_{\textmd{Q}}b\big)\big]
\\\label{41}
&\times&\big[6r^{3}(b+1)(2b-1)(r-\nu)\big]^{-1}.
\end{eqnarray}

In the context of \textit{WH} studies, the redshift function plays a
crucial role in determining the gravitational redshift experienced
by signals traveling through the \textit{WH}. Physically, it
represents how the gravitational potential changes along the radial
coordinate of the \textit{WH} throat and its surrounding geometry.
When the redshift function is constant, it implies that there is no
gravitational redshift experienced by signals passing through the
\textit{WH}. This is often a simplifying assumption in many
\textit{WH} models to avoid singularities at the throat and to ease
the analysis of the spacetime geometry.
\begin{itemize}
\item
Physical Meaning of a Constant Redshift Function
\end{itemize}
A constant redshift function ensures that the \textit{WH} spacetime
does not have an event horizon. Event horizons are typically
associated with divergent redshift functions (as in the case of
black holes), where signals cannot escape beyond a certain region.
With constant redshift, signals can travel freely through the
\textit{WH}, making it a traversable \textit{WH}. Since redshift
function is related to the gravitational time dilation, a constant
value indicates that there is no differential time dilation across
the \textit{WH} geometry. This can be important in constructing
traversable \textit{WHs} that are stable for interstellar travel, as
travelers moving through the \textit{WH} would not experience time
distortion.
\begin{itemize}
\item
Impact on the Modified Field Equations
\end{itemize}
When the redshift function is constant, it simplifies the Einstein
field equations or any modified versions of these equations that
govern the \textit{WH} spacetime. The modified Einstein field
equations often contain derivatives of redsgift function. When
redshift function is constant, terms involving derivatives of
redshift vanish, significantly reducing the complexity of the
equations. In theories of modified gravity, a constant redshift
function might change the way curvature terms interact with the
geometry, affecting the required modifications to the field
equations. The lack of a varying potential could reduce the
complexity of the additional terms introduced by these theories,S
leading to more manageable conditions for finding solutions. Thus, a
constant redshift function in \textit{WH} studies simplifies the
field equations by eliminating time dilation effects and reducing
the complexity of the spacetime metric. It affects the physical
characteristics of the \textit{WH}, such as the absence of an event
horizon and simplified matter-energy conditions needed to sustain
the \textit{WH} structure.

Astrophysical objects are composed of a variety of materials.
Determining the kind of substance (ordinary or exotic) that is
contained in celestial objects is essential. In order to comprehend
the nature of matter in the cosmic objects, we look into certain
inequalities, known as \textit{ECs}. These constraints are useful in
confirming the viability of \textit{WHs}. The \textit{ECs} for the
anisotropic configuration are expressed as null energy bound $(0\leq
p_{r}+\rho, \quad 0\leq p_{\bot}+\rho)$, dominant energy bound
$(0\leq \rho\pm p_{r}, \quad 0\leq \rho\pm p_{\bot})$, weak energy
bound $(0\leq p_{r}+\rho,\quad 0\leq p_{\bot}+\rho, \quad 0\leq
\rho)$ and strong energy bound $(0\leq p_{r}+\rho, \quad 0\leq
p_{\bot}+\rho, \quad 0\leq p_{r}+2p_{\bot}+\rho)$. The existence of
traversable \textit{WH} geometries and other hypothetical objects in
spacetime requires a comprehension of these \textit{ECs}. A feasible
\textit{WH} structure must diverge from these conditions.

\subsection{Viable $f(\textmd{Q},\textmd{T})$ Models}

Here, we examine the effects of several models of
$f(\textmd{Q},\textmd{T})$ theory on the geometry of \textit{WH}.
Our work attempts to reveal obscure astrophysical and theoretical
cosmological insights. The correction terms of this modified
gravitational theory could produce insightful findings. The
existence of feasible cosmic geometries could be significantly
impacted by these modified terms. Therefore, exploring
$f(\textmd{Q},\textmd{T})$ becomes crucial in identifying
hypothetical objects. We examine three different models of
$f(\textmd{Q},\textmd{T})$ in the following subsections.

\subsection*{Model 1}

In this context, we examine a power-law $f(\textmd{Q},\textmd{T})$
model with arbitrary constant $a_{1}$ as
\begin{equation}\label{41a}
f(\textmd{Q},\textmd{T})=\textmd{Q}+a_{1} \textmd{Q}^{2}+b
\textmd{T}.
\end{equation}
This model has important cosmological implications, particularly in
explaining early universe dynamics and late-time cosmic
acceleration. It provides a framework for exploring alternatives to
dark energy and possibly modifying gravitational wave predictions,
making it a significant candidate for describing the cosmic
evolution. The quadratic correction $a_{1} \textmd{Q}^{2}$
introduces non-linearities in the gravitational field equations,
which could have significant cosmological implications. Such
quadratic terms are known to affect the early universe dynamics and
they may lead to the possibility of a cosmological bounce, avoiding
the singularity problem of the Big Bang. This term could also
provide corrections to the late-time evolution of the universe,
influencing cosmic acceleration or mimicking dark energy effects.
The coupling of the trace of the energy-momentum tensor with gravity
represents a direct interaction between matter and geometry. This
interaction implies that the energy content of the universe could
influence the gravitational field in ways different from
\textit{GTR}. The term $b \textmd{T}$ is expected to impact the
equation of state of cosmological fluids, possibly leading to
modifications in how matter and radiation evolve in the universe. It
could offer explanations for cosmic acceleration without the need
for dark energy by introducing effective pressure contributions
arising from the coupling term. The combination of non-metricity and
matter-geometry coupling introduces new dynamics in the cosmic
evolution. Depending on the values of the parameters, this model
could lead to deviations from the standard cosmological model,
influencing the structure formation and the overall dynamics of
cosmic expansion. We consider the model parameters values as $a_{1}=
0.7$ and $b=0.9$ to examine the viable \textit{WH} solutions.

The field equations corresponding to this model are provided in
Appendix \textbf{A}. We now look at an interesting case with a
redshift function that is constant. The derivation of exact
\textit{WH} solutions is made possible by this simplification, which
also significantly simplifies computations. In order to investigate
the feasible \textit{WH} geometry, we take into account the various
shape functions in the following scenarios. In all the cases, we
take \textit{WH} throat at 0.5, i.e., $r_{0}=0.5$ for our
convenience. It is important to mention that the all Morris-Thorne
conditions are satisfied at $r_{0}=0.5$.

\subsubsection*{Case 1: $\nu(r)=\frac{r_{0}^{2}}{r}$ }

We consider the specific choice of shape function as
$\nu(r)=\frac{r_{0}^{2}}{r}$ \cite{37a}. The field equations
corresponding to this case turn out to be
\begin{eqnarray}\nonumber
\rho&=&\frac{2a_{1}r_{0}^{6}((23b-27)
r_{0}^{2}+14(3-2b)r^{2})-(2b-3)r^{4}(r_{0}^{3}-r_{0}
r^{2})^{2}}{3r^{8}(2b^{2}+b-1)(r_{0}^{2}-r^{2})^{2}},
\\\nonumber
p_{r}&=&\frac{2a_{1}r_{0}^{6}((b+15)r_{0}^{2}+2(2b-15)r^{2})
-(10b-3)r^{4}(r_{0}^{3}-r_{0}r^{2})^{2}}{3r^{8}(2b^{2}+b-1)
(r_{0}^{2}-r^{2})^{2}},
\\\nonumber
p_{\bot}&=&\frac{(2b-3)r^{4}(r_{0}^{3}-gr^{2})^{2}+2a_{1}r_{0}^{6}
((25b+3)r_{0}^{2}-2(22
b+3)r^{2})}{3(2b^{2}+b-1)r^{8}(r_{0}^{2}-r^{2})^{2}}.
\end{eqnarray}
The graphical behavior of the \textit{ECs} for different values of
$a_{1}$ and $b$ is shown in Figure \textbf{1}. The behavior of
\textit{ECs} is examined in the upper panel for small values of the
model parameters. The matter contents $\rho$, $\rho-p_{r}$ and
$\rho-p_{\bot}$ in the left plot show negative behavior which
violates the dominant and weak \textit{ECs}, whereas all
\textit{ECs} are violated in the right plot. It is clear from the
lower panel that for both large positive and negative parametric
values, the dominant \textit{EC} is violated. These graphs show that
the fluid variables violate the \textit{ECs}, which provides the
feasible traversable \textit{WH} structures in this case.
\begin{figure}
\epsfig{file=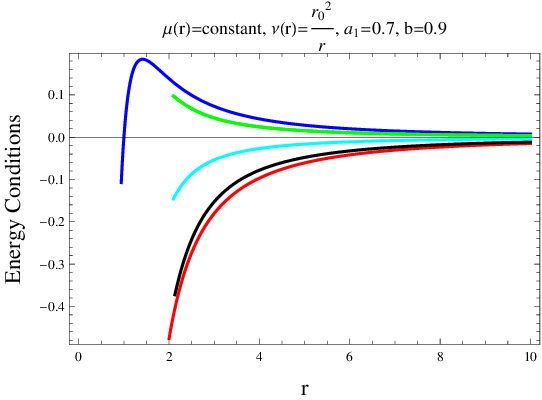,width=.5\linewidth}
\epsfig{file=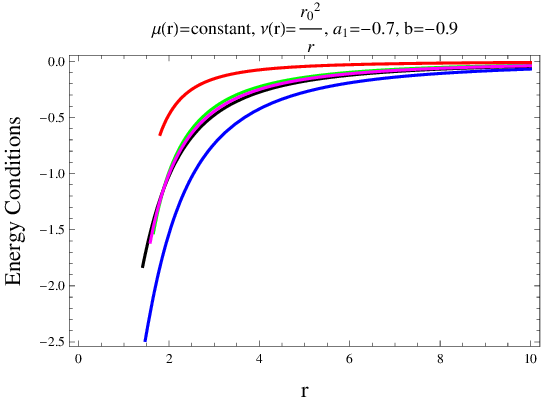,width=.5\linewidth}
\epsfig{file=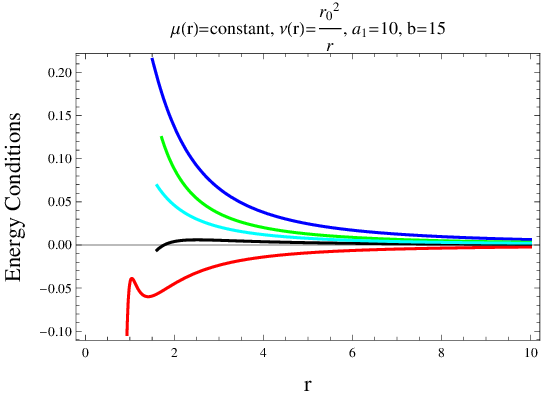,width=.5\linewidth}
\epsfig{file=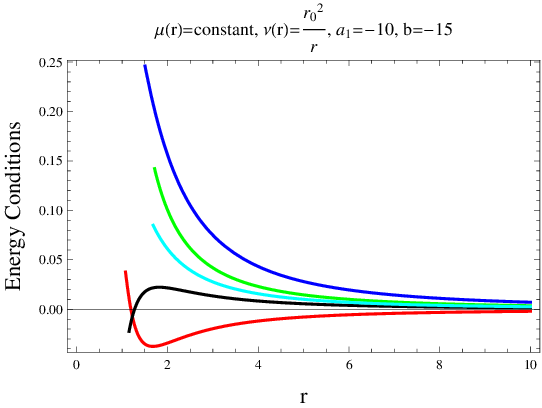,width=.5\linewidth}\caption{Graphs of
$\rho+p_{r}$ (blue), $\rho+p_{\bot}$ (green), $\rho-p_{r}$ (red),
$\rho-p_{\bot}$ (black) and $\rho$ (cyan) corresponding to Model 1
(Case 1) for different parametric values.}
\end{figure}

\subsubsection*{Case 2: $\nu(r)=re^{r_{0}-r}$}

The field equations with respect to $\nu(r)=re^{r_{0}-r}$ \cite{37a}
and constant redshift are
\begin{eqnarray}\nonumber
\rho&=&\frac{1}{6r^{2}(2b^{2}+b-1)
(e^{r_{0}}-e^{r})^{2}}\big[e^{{r_{0}-2r}}\big(a_{1}
e^{2r_{0}}((19b-21)e^{r_{0}}+(36-24
\\\nonumber
&\times&b)e^{r})-2e^{r}(2b-3)(r-1)(e^{r_{0}}-e^{r})^{2}\big)\big],
\\\nonumber
p_{r}&=&\frac{1}{6r^{2}(2b^{2}+b-1)(e^{r_{0}}-e^r)^{2}}\big[e^{r_{0}-2r}\big(a_{1}e^{{2
r_{0}}}((5b+9)e^{r_{0}}-24e^{r})-2e^{r}
\\\nonumber
&\times&(b(4r+2)-3)(e^{r_{0}}-e^{r})^{2}\big)\big],
\\\nonumber
p_{\bot}&=&\frac{1}{6r^{2}(2b^{2}+b-1)
(e^{r_{0}}-e^{r})^{2}}\big[e^{r_{0}-2r}\big(a _{1}e^{2r_{0}}((17b+3)
e^{r_{0}}-6(6b+1)
\\\nonumber
&\times& e^{r})-e^{r}(2b(r-4)+3r)(e^{r_{0}}-e^{r})^{2}\big)\big].
\end{eqnarray}
The graphical representation of the \textit{ECs} for various values
of $a_{1}$ and $b$ corresponding to case 2 is shown in Figure
\textbf{2}. It can be seen in the upper and lower panels that we
obtain the identical behavior as found in case 1., i.e., the matter
contents $(\rho-p_{r},\rho-p_{\bot},\rho)$ behave negatively in the
upper panel and dominant \textit{EC} is violated in the lower panel.
Thus, we find that the viable traversable \textit{WH} structures
exist for both positive and negative values of the model parameters.
\begin{figure}
\epsfig{file=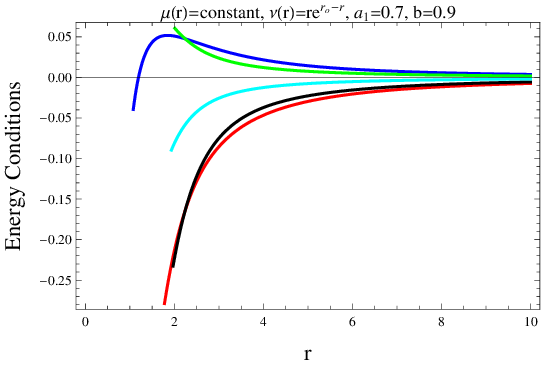,width=.5\linewidth}
\epsfig{file=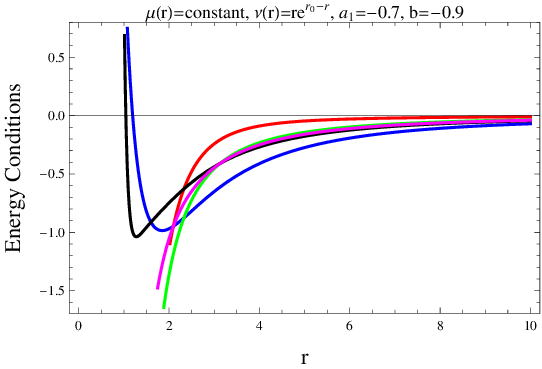,width=.5\linewidth}
\epsfig{file=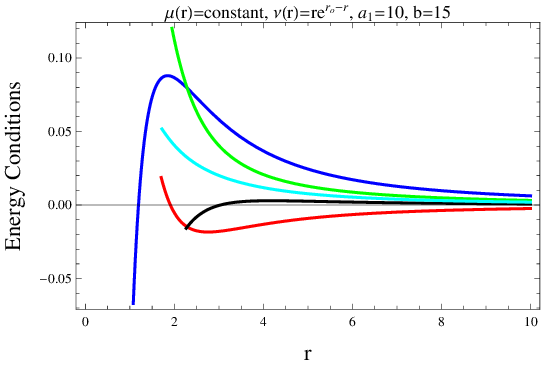,width=.5\linewidth}
\epsfig{file=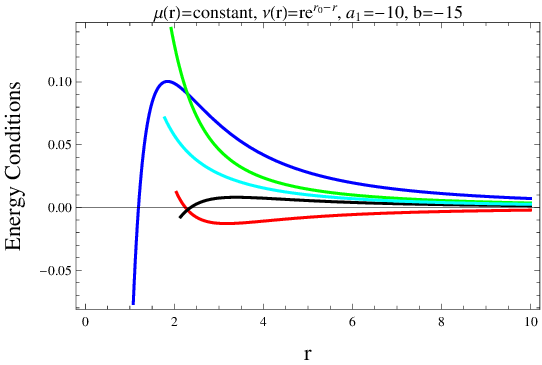,width=.5\linewidth}\caption{Graphs of
$\rho+p_{r}$ (blue), $\rho+p_{\bot}$ (green), $\rho-p_{r}$ (red),
$\rho-p_{\bot}$ (black) and $\rho$ (cyan) corresponding to Model 1
(Case 2) for different parametric values.}
\end{figure}

\subsubsection*{Case 3: $\nu(r)=r_{0}e^{1-\frac{r}{r_{0}}}$}

The corresponding field equations for
$\nu(r)=r_{0}e^{1-\frac{r}{r_{0}}}$ \cite{37b} are
\begin{eqnarray}\nonumber
\rho&=&\frac{1}{6(2b^{2}+b-1)r^{6}(er_{0}-r
e^{r/r_{0}})^2}\big[e^{1-\frac{2r}{r_{0}}}\big\{e^{2}a_{1}
r_{0}\big(er_{0}(3(9b-11)r_{0}^{2}+2
\\\nonumber
&\times&(19b-21)r_{0}r+(19b-21)r^{2})-4(2b-3)r
e^{r/r_{0}}(4r_{0}^{2}+6r_{0}r+3r^{2})\big)
\\\nonumber
&-&2(2b-3)r^{4}e^{r/r_{0}}(er_{0}-re^{r/r_{0}})^{2}\big\}\big],
\\\nonumber
p_{r}&=&\frac{1}{6(2b^{2}+b-1)r^{6}
(er_{0}-re^{r/r_{0}})^{2}}\big[e^{1-\frac{2r}{r_{0}}}\big\{e^{2}a
_{1}r_{0}\big(er_{0}(-3(b-7)r_{0}^{2}+2
\\\nonumber
&\times&(5b+9)r_{0}r+(5b+9)r^{2})-4re^{r/r_{0}}((9-2b)r_{0}^{2}+12r_{0}r+6r^{2})\big)-2
\\\nonumber
&\times&r^{3}e^{r/r_{0}}(er_{0}-re^{r/r_{0}})^{2}((6b-3)r_{0}+4b
r)\big\}\big],
\\\nonumber
p_{\bot}&=&\frac{1}{6(2b^{2}+b-1)r^{6}(er_{0}-r
e^{r/r_{0}})^{2}}\big[e^{1-\frac{2r}{r_{0}}}\big\{e^{2}a_{1}r_{0}
\big(er_{0}((33b+3)r_{0}^{2}+2
\\\nonumber
&\times&(17b+3)r_{0}r+(17b+3)r^{2})-2re^{r/r_{0}}((26b+3)r_{0}^{2}+6r_{0}(6br+r)
\\\nonumber
&+&3(6b+1)r^{2})\big)-r^{3}e^{r/r_{0}}(er_{0}-re^{r/r_{0}})^{2}((3-6
b)r_{0}+(2 b+3)r)\big\}\big].
\end{eqnarray}
Figure \textbf{3} determines that the matter components $(\rho,
\rho\pm p_{r}, \rho\pm p_{\bot})$ exhibit negativity across all
parametric values. This deviation from \textit{ECs} suggests the
presence of exotic matter, providing justification for the
feasibility of a traversable \textit{WH} geometry in this
gravitational model.
\begin{figure}
\epsfig{file=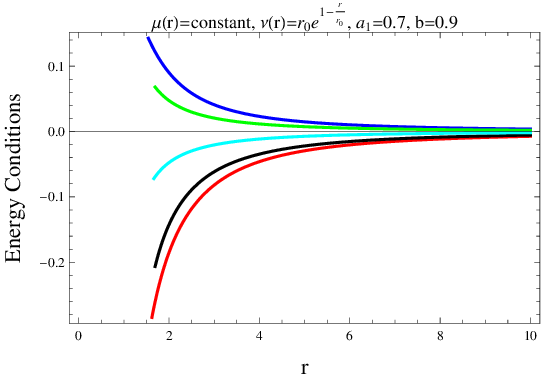,width=.5\linewidth}
\epsfig{file=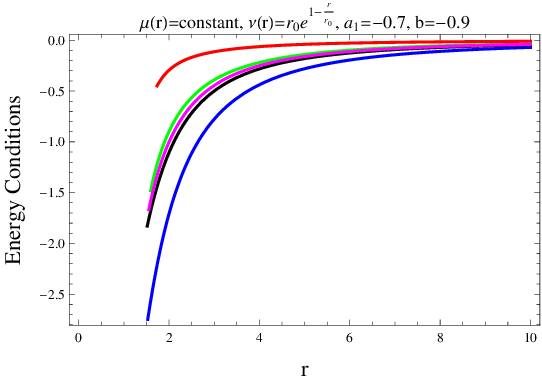,width=.5\linewidth}
\epsfig{file=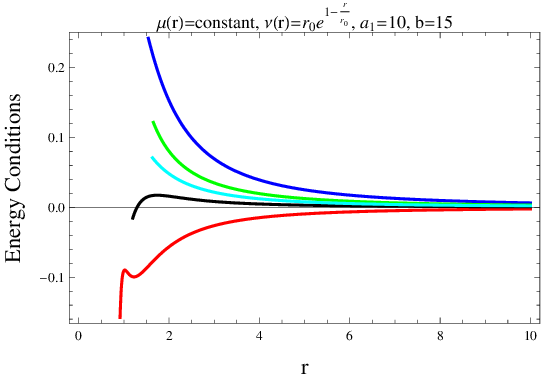,width=.5\linewidth}
\epsfig{file=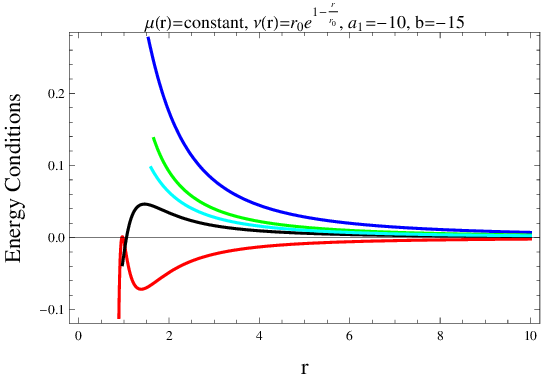,width=.5\linewidth}\caption{Graphs of
$\rho+p_{r}$ (blue), $\rho+p_{\bot}$ (green), $\rho-p_{r}$ (red),
$\rho-p_{\bot}$ (black) and $\rho$ (cyan) corresponding to Model 1
(Case 3) for different parametric values.}
\end{figure}

\subsubsection*{Case 4: $\nu(r)=r\frac{\ln{r+1}}{r_{0}+1}$}

In this specific case \cite{37c}, we obtain the field equations as
\begin{eqnarray}\nonumber
\rho&=&\frac{1}{6(2b^{2}+b-1)(r_{0}+1)^{4}(r+1)^{2}
(r-\frac{r\ln(r+1)}{r_{0}+1})^{2}}\big[a_{1}\ln(r+1)(\ln(r+1)(8
\\\nonumber
&\times&br_{0}+(12-8b)\ln(r+1)+19b-12r_{0}-21)-8(2b-3)(r_{0}+1))+2(2
\\\nonumber
&\times&b-3)(r_{0}+1)(r+1)(r+(r+1)\ln(r+1))(r_{0}-\ln(r+1)+1)^{2}\big],
\\\nonumber
p_{r}&=&\frac{1}{6(2b^{2}+b-1)(g+1)^{2}r^{2}(r+1)^{2}(r_{0}
-\ln(r+1)+1)^{2}}\big[a_{1}\ln(r+1)(\ln
\\\nonumber
&\times&(r+1)(-8br_{0}+4(2b-3)\ln(r+1)+5b+12 r_{0}+9)-4(2b+3)(r_{0}
\\\nonumber
&+&1))-2(r_{0}+1)(r+1)((2b-3)(r+1)\ln(r+1)-4br)(r_{0}-\ln(r+1)
\\\nonumber
&+&1)^{2}\big],
\\\nonumber
p_{\bot}&=&\frac{1}{6(2b^{2}+b-1)(r_{0}+1)^{2} r^{2}(r+1)^{2}
(r_{0}-\ln(r+1)+1)^{2}}\big[(r_{0}+1)(r
\\\nonumber
&+&1)((2b+3)r+8 b(r+1)\ln(r+1))(r_{0}-\ln(r+1)+1)^{2}-a_{1}\ln (r
\\\nonumber
&+&1)(\ln(r+1)(-16br_{0}+16b \ln (r+1)-17
b-3)+2(10b+3)(r_{0}
\\\nonumber
&+&1))\big].
\end{eqnarray}
Figure \textbf{4} violates the \textit{ECs} which manifests the
existence of viable traversable \textit{WH} geometry in this case.
\begin{figure}
\epsfig{file=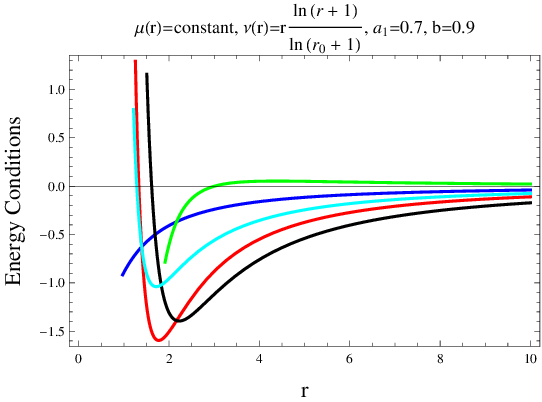,width=.5\linewidth}
\epsfig{file=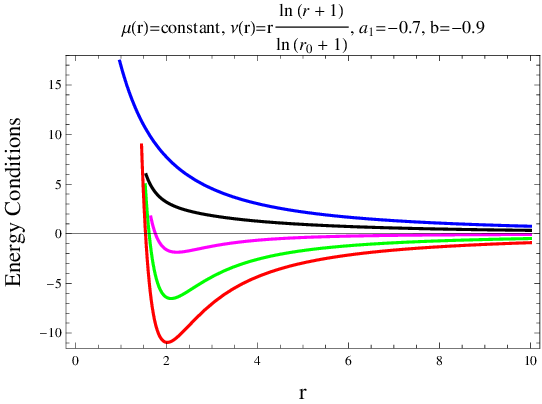,width=.5\linewidth}
\epsfig{file=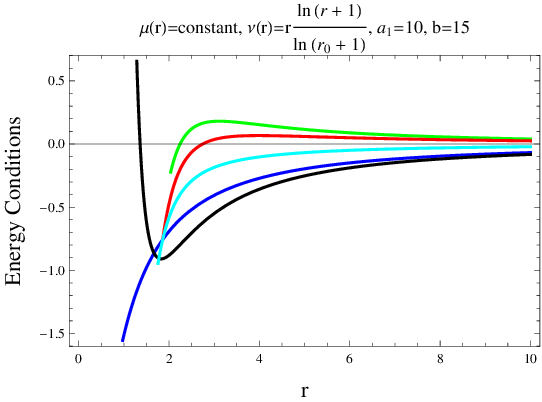,width=.5\linewidth}
\epsfig{file=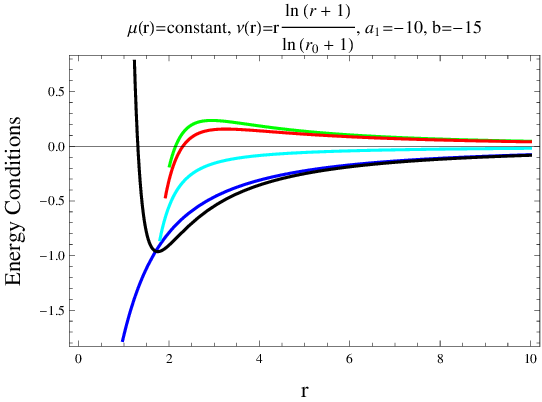,width=.5\linewidth}\caption{Graphs of
$\rho+p_{r}$ (blue), $\rho+p_{\bot}$ (green), $\rho-p_{r}$ (red),
$\rho-p_{\bot}$ (black) and $\rho$ (cyan) corresponding to Model 1
(Case 4) for different parametric values.}
\end{figure}

\subsection*{Model 2}

Here, we use another model with constant $a_{2}$ as \cite{37a}
\begin{equation}\label{41b}
f(\textmd{Q},\textmd{T})=\textmd{Q}+\frac{a_{2}}{\textmd{Q}}+b
\textmd{T},
\end{equation}
This model has notable cosmological significance, particularly in
addressing key issues like the early universe evolution, late-time
cosmic acceleration and potential avoidance of singularities. Its
impact on the early universe and the late-time accelerated expansion
make it a candidate for explaining phenomena that the standard
$\Lambda$CDM model might not fully capture. Additionally,
observational signatures such as deviations in gravitational wave
propagation or cosmic structure growth could help test the viability
of this model. The term $\frac{a_{2}}{\textmd{Q}}$ introduces a
non-linear modification to the non-metricity, which can have
significant implications for the cosmological dynamics, particularly
in the early universe and at large scales. The presence of this
inverse term suggests that when $\textmd{Q}$ becomes small (as it
might near a bounce or at high-curvature regions), the dynamics
become strongly modified. This term could contribute to avoiding
cosmological singularities, leading to bouncing solutions or
early-time inflation. The presence of the inverse term could also
help in avoiding future singularities such as the Big Rip or other
catastrophic fates, by altering the dynamics when $\textmd{Q}$
becomes small. This could stabilize the cosmological evolution and
lead to a more regular, non-singular universe. We consider $a_{2}=
0.8$ and $b=0.9$ to analyze the viable \textit{WH} structures.

The modified field equations for this model are given in Appendix
\textbf{B}. In this model, the graphical representations
corresponding to cases \textbf{1-3} indicate the non-consistent
behavior of the fluid parameters for both positive as well as
negative values of the model parameters. This suggests that this
model does not support the geometry of \textit{WH} corresponding to
specific shape functions given in the cases \textbf{1-3}.

\subsubsection*{Case 4: $\nu(r)=r\frac{\ln{r+1}}{r_{0}+1}$}

This case yields the field equations in the following form
\begin{eqnarray}\nonumber
\rho&=&-\frac{1}{6r^{5}(2b^{2}+b-1)\ln
^{2}(r+1)}\big[(r_{0}+1)(r+1)\big\{r^{5}(r_{0}-\ln(r+1)+1)
\\\nonumber
&\times&\big(a_{2}
(\ln(r+1)(r(20br_{0}+13b-30r_{0}-27)+6(2b-3)(r_{0}+1)-2 (2b
\\\nonumber
&-&3)(5r+3) \ln(r+1))-2(2b-3)(r_{0}+1)r)-3a_{2}(b+1)r\ln (r+1)\big)
\\\nonumber
&-&(\frac{1}{(r_{0}+1)^{2}(r+1)^{2}}(2(2b-3)r^{3}\ln^{2}(r+1)(r+(r+1)\ln
(r+1))))\big\}\big],
\\\nonumber
p_{r}&=&-\frac{1}{6r^{2}(2b^{2}+b-1)(r_{0}+1)(r+1)
\ln^{2}(r+1)}\big[2\ln^{2}(r+1)((2b-3)
\\\nonumber
&\times&(r+1)\ln(r+1)-4br)-(r_{0}+1)^2r^{2}(r+1)^{2}({r_{0}}-\ln(r+1)+1)
\\\nonumber
&\times&(a_{2}(\ln(r+1)(r(5b(4
r_{0}+5)-30r_{0}-33)+6(2b-3)(r_{0}+1)-2(2b
\\\nonumber
&-&3)(5 r+3)\ln(r+1))-4(4b-3)
(r_{0}+1)r)-3a_{2}(b+1)r\ln(r+1))\big],
\\\nonumber
p_{\bot}&=&-\frac{1}{6r^{2}(2b^{2}+b-1)(r_{0}+1)(r+1)\ln
^{2}(r+1)}\big[(r_{0}+1)^{2}r^{2}(r+1)^{2}(r_{0}
\\\nonumber
&-&\ln(r+1)+1)(a_{2}(b\ln (r+1)(8r_{0}(5r+3)+41r-8(5r+3)\ln (r
\\\nonumber
&+&1)+24)-(14b-3)(r_{0}+1)r)+3a_{2}(b+1)r\ln(r +1))+\ln^{2}(r
\\\nonumber
&+&1)(-(2b+3) r-8b(r+1)\ln(r+1))\big].
\end{eqnarray}
In Figure \textbf{8}, the upper panels illustrate the negative
trends of $\rho$, $\rho+p_{r}$ and $\rho-p_{\bot}$, indicating the
violation of the \textit{ECs}. Similarly, all \textit{ECs} are
observed to be violated in the lower panel. These graphical
representations indicate that the fluid parameters violate the
\textit{ECs} across positive and negative values of model
parameters. This suggests the presence of viable traversable
\textit{WH} structures in this framework.
\begin{figure}
\epsfig{file=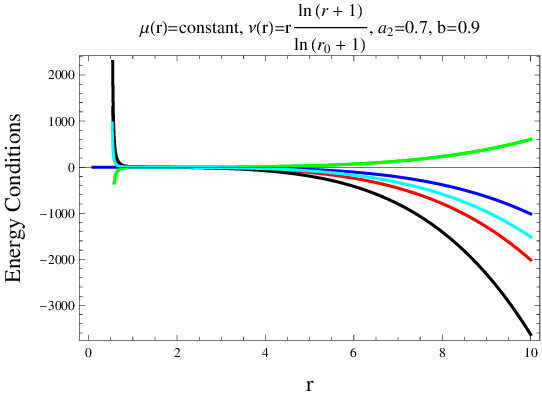,width=.5\linewidth}
\epsfig{file=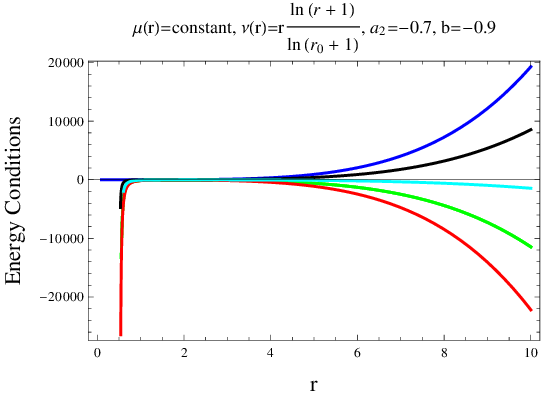,width=.5\linewidth}
\epsfig{file=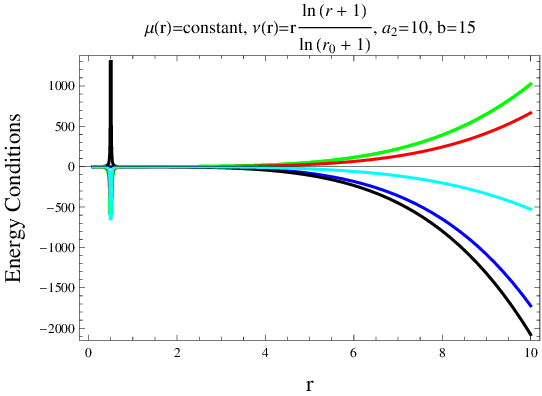,width=.5\linewidth}
\epsfig{file=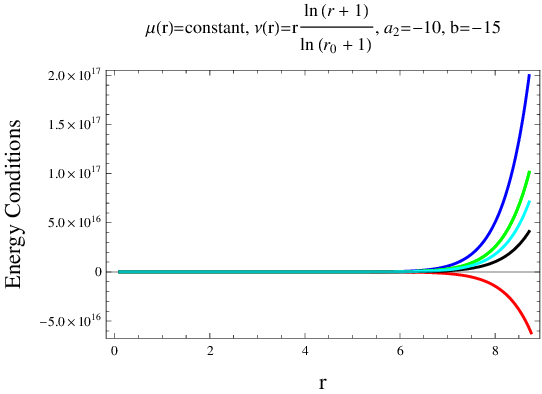,width=.5\linewidth}\caption{Graphs of
$\rho+p_{r}$ (blue), $\rho+p_{\bot}$ (green), $\rho-p_{r}$ (red),
$\rho-p_{\bot}$ (black) and $\rho$ (cyan) corresponding to Model
2(Case 4) for different parametric values.}
\end{figure}

\subsection*{Model 3}

Finally, we consider \cite{37a}
\begin{equation}\label{41c}
f(\textmd{Q},\textmd{T})=\textmd{Q}e^{\frac{a_{3}}{\textmd{Q}}}+b
\textmd{T}.
\end{equation}
This mode has potential cosmological significance. Its non-linear
modification to gravity through the non-metricity scalar can provide
rich dynamics that go beyond \textit{GTR}. Additionally, the
coupling to the trace of energy-momentum tensor lead to novel
effects in the matter-gravity interaction. This model could be
relevant for addressing key cosmological issues like the nature of
dark energy, the origin of cosmic inflation, or the behavior of
gravity at large scales. The term
$\textmd{Q}e^{\frac{a_{3}}{\textmd{Q}}}$ introduces a non-linear
interaction of $\textmd{Q}$, which might give rise to interesting
cosmological effects. The exponential function suggests that the
theory introduces corrections to \textit{GTR} that become
significant in regimes where $Q$ is small, i.e., in the early or
late universe when the geometry is evolving rapidly. The exponential
term could play a role in the early universe. This might result in
inflationary dynamics or provide an alternative to scalar field
inflation models by driving an accelerated expansion. Depending on
the values of the model might admit solutions that describe
accelerating expansion, either in the early universe (inflation) or
at late times (cosmic acceleration). For viable \textit{WH}
geometry, we assume the constant values as $a_{3}= 0.8$ and $b=0.9$.

The corresponding field equations for this model are bestowed in
Appendix \textbf{C}. In this model, the graphical representation of
the matter components $(\rho, \rho\pm p_{r}, \rho\pm p_{\bot})$
indicate that \textit{WH} does not exist corresponding to this model
for all cases. Hence, it is found that this gravity model does not
support the existence of a viable traversable \textit{WH} structures
for all cases. Thus, we conclude that this
$f(\textmd{Q},\textmd{T})$ model is not suitable for \textit{WH}
solution.

\section{Final Remarks}

The exploration of \textit{WH} structures holds paramount importance
in the realm of astrophysics. Exotic matter is essential for
physically viable \textit{WH} geometries. Modified theories of
gravity have drawn a lot of attention from the scientific community
in recent decades as a potential replacement for \textit{GTR}. These
modified theories examine the viability of traversable \textit{WH}
geometries by the correction term ability to violate \textit{ECs}
and guarantee the existence of viable \textit{WH} structures.
Modified theories are not just important for theoretical conjecture,
but they are essential for revealing the physical characteristics of
celestial objects. In order to investigate the feasible \textit{WH}
configurations in the context of $f(\textmd{Q},\textmd{T})$ theory,
we have considered the Morris-Throne spacetime with anisotropic
matter configuration. To get the explicit expressions of the
pressure and energy density components, we have assumed the various
models of this modified theory. The viability of \textit{WH}
structures was then examined using \textit{ECs} with various shape
functions and a constant redshift function.

In this manuscript, we have investigated the feasibility of
traversable \textit{WH} solutions in the framework of
$f(\textmd{Q},\textmd{T})$ theory. Our primary aim is to examine the
theoretical aspects of \textit{WH} solutions and their behavior
under modified terms. Traversable \textit{WHs} have garnered
significant attention in theoretical physics due to their potential
to connect different regions of spacetime. Identifying viable and
stable traversable \textit{WH} solutions in
$f(\textmd{Q},\textmd{T})$ theory is not just a theoretical exercise
rather, it holds profound implications for our understanding of the
fundamental laws governing the universe. Firstly, our findings
suggest that traversable \textit{WHs} can exist in modified gravity
theories such as $f(\textmd{Q},\textmd{T})$ theory. This challenges
the concept that such structures are only theoretically feasible in
the framework of \textit{GTR}. By demonstrating their existence in
this modified framework, we have expanded the scope for studying and
investigating these intriguing phenomena with a wider range of
theoretical frameworks than previously considered. Thus, our
research advances the understanding of traversable \textit{WHs} in a
modified framework and lays a theoretical foundation for further
exploration.

Traversable \textit{WHs} represent a captivating subject in
theoretical physics, particularly in the framework of
$f(\textmd{Q},\textmd{T})$ theory. The exploration of \textit{WHs}
holds significant relevance and theoretical significance. Their
existence introduces the intriguing prospect of time travel by
enabling the formation of closed timelike curves. In the context of
$f(\textmd{Q},\textmd{T})$ gravitational theory, the presence of
traversable \textit{WHs} challenges conventional notions of
spacetime geometry. It implies that deviations from the standard
Einstein-Hilbert action can yield exotic structures like
\textit{WHs}, thereby opening up new avenues for investigating the
foundational principles of physics. The analysis of traversable
\textit{WHs} offers valuable insights into the impact of modified
gravity on the large-scale structure of the universe. This
contributes to refining our models of cosmic evolution and
understanding the eventual fate of the universe. Moreover, the
existence of viable traversable \textit{WHs} in
$f(\textmd{Q},\textmd{T})$ theory facilitates a deeper exploration
of the \textit{ECs} governing spacetime and their compatibility with
exotic matter. Consequently, the practical implications of
traversable \textit{WHs} in $f(\textmd{Q},\textmd{T})$ theory have
the potential to revolutionize both space travel and cosmology,
while their theoretical ramifications reshape our comprehension of
fundamental physics, encompassing gravity, spacetime and the essence
of the cosmos.

For both positive and negative values of the model parameters, our
analysis in the first model shows the violation of \textit{ECs}
which indicates the presence of exotic matter at the \textit{WH}
throat. The dominate \textit{EC} is found to be violated for all
values of $a_{1}$ and $b$ for the cases 1 and 2. This gives a
traversable \textit{WH} structure in this scenario (Figures
\textbf{1} and \textbf{2}). For the cases \textbf{3} and \textbf{4},
we also obtain the feasible traversable \textit{WH} geometry
(Figures \textbf{3} and \textbf{4}). The traversable \textit{WH}
geometry in the model 2 is only obtained for case 4 as the existence
of exotic matter at the \textit{WH} throat is ensured by the
negative trends of fluid parameters, which indicate a violation of
the \textit{ECs} (Figures \textbf{5}). We have noted that the
\textit{WH} geometry for model 2 is not supported by the behavior of
matter variables corresponding to cases 1-3. Additionally, it is
discovered that the \textit{WH} geometry corresponding to all cases
is not supported by model \textbf{3}. Thus, the only model that
supports the presence of a feasible traversable \textit{WH}
structure is $f(\textmd{Q},\textmd{T})=\textmd{Q}+a_{1}
\textmd{Q}^{2}+b \textmd{T}$.

The geometric features of \textit{WHs} in $f(\textmd{R})$ gravity
were studied by Lobo and Oliveira in \cite{42}, who found that there
are no feasible \textit{WH} solutions in vacuum. In $f(\textmd{R})$
theory, Fayyaz and Shamir \cite{43} investigated physically
realistic traversable \textit{WH} configurations using the Karmarkar
condition. They found viable solutions with minimum exotic matter.
In order to investigate \textit{WHs} in $f(\textmd{R},\textmd{T})$
theory, Sharif and Fatima \cite{44} found \textit{WH} solutions for
minimum radius. According to Banerjee et al \cite{45}, the power-law
model $f(\textmd{Q})=\textmd{Q}+a_{1} \textmd{Q}^{2}$ does not have
any \textit{WH} solution. It is worthwhile to mention here that we
have obtained viable \textit{WH} solutions in
$f(\textmd{Q},\textmd{T})$ theory corresponding to the functional
form of $f(\textmd{Q})=\textmd{Q}+a_{1}\textmd{Q}^{2}$ in which
Banerjee et al \cite{45} found that the viable WH solution does not
exist corresponding to this functional form.

The observational evidence for \textit{WHs} presents a promising
avenue for advancing our comprehension of the cosmos. To validate
\textit{WH} solutions, future studies and experiments could consider
the following directions.
\begin{itemize}
\item
Gravitational Lensing
\\\\
Investigate the detection of distinct gravitational lensing
signatures linked to traversable \textit{WHs}. Analyzing lensing
effects on background light sources might unveil characteristic
patterns indicative of \textit{WH} presence.
\item
Multi-Messenger Astronomy
\\\\
Analyze gravitational wave observations with various forms of
electromagnetic radiation such as gamma-ray bursts or neutrino
detections. Coordinated multi-messenger observations could furnish a
comprehensive dataset for identifying viable \textit{WH} events.
\item
Cosmic Microwave Background
\\\\
Examine anomalies in the cosmic microwave background that may be
attributed to the influence of nearby \textit{WHs}. Deviations in
temperature or polarization patterns could serve as indirect
indicators of these exotic structures.
\item
Astrophysical Disk Dynamics
\\\\
Investigate the dynamics of accretion disks around black holes to
detect deviations that could be linked to the presence of
\textit{WHs}. Variations in disk behavior might offer indirect
evidence of nearby traversable \textit{WHs}.
\item
High-Energy Particle Colliders
\\\\
Explore the viability of probing the microscopic scale for evidence
of exotic matter or energy consistent with the theoretical
requirements for stabilizing and sustaining \textit{WHs}. Particle
accelerator experiments may provide insights into the fundamental
physics associated with these structures.
\item
Time-Delay Observations
\\\\
Conduct precise timing observations of distant astrophysical events,
such as gamma-ray bursts, and look for unexpected time delays that
could be attributed to gravitational effects associated with
traversable \textit{WHs}.
\end{itemize}
These proposed avenues for future research aim to inspire the
scientific community to explore diverse observational methods,
fostering a viable approach to validating the existence of
\textit{WHs} and advancing our understanding of the fundamental
nature of the universe.

\section*{Appendix A: Field Equations (Model 1)}

Using Eqs.(\ref{39})-(\ref{41}), we obtain the following field
equations corresponding to model 1 as
\begin{eqnarray}\nonumber
\rho&=&\frac{1}{6r^{6}(2b^{2}+b-1)(r-\nu)^{2}}
\bigg[r^{4}(r-\nu)^{2}\big(10b
r(r-\nu)\mu''-5b\mu'(r(\nu'-4)
\\\nonumber
&+&3\nu)+10br(r-\nu)\mu'^{2}+(4b-6)\nu'\big)+a_{1}
\big\{-20br^{5}\mu'\nu'(2r\mu'+\nu')-2
\\\nonumber
&\times&r^{3}\nu\big(20br^{3}\mu'^{3}+2\nu'(5br^{2}
\mu''+(4b-6)\nu')-60
br^{2}\mu'^{2}\nu'+r\mu'\big(10br(4r
\\\nonumber
&\times&\mu
''+\nu'')-15b\nu'^{2}-2(7b+12)\nu'\big)\big)+r^{2}\nu^{2}
\big\{120br^{3}\mu'^{3}+8\nu'(5br^{2}\mu''
\\\nonumber
&+&6b-9)+12r^{2}\mu'^{2}(-10b\nu'+b+1)+4r((b+6)r\mu''+(3-2
b)\nu'')+2
\\\nonumber
&\times&r\mu'\big(20br(6r\mu''+\nu'')-5b\nu'^{2}-6(2b+7)\nu'-4
(b+6)\big)+(11 b-9)\nu'^{2}\big\}
\\\nonumber
&-&2r\nu^{3}\big(60br^{3}\mu'^{3}+\nu'(10br^{2}\mu''+19b-21)+4
(b+6)r^{2}\mu''+4r^{2}\mu'^{2}(-5b
\\\nonumber
&\times&\nu'+3 b+3)+r\mu'(10br(12r\mu''+\nu'')+2(b-9)\nu'+3
(b-14))+2(3
\\\nonumber
&-&2b)r\nu''+8(2b-3)\big)+\nu^{4}\big(2r
\big(\mu'(2r\mu'(10br\mu'+3
b+3)+7 b-18)
\\\nonumber
&+&2r\mu''(20br\mu'+b+6)\big)+27b-33\big)\big\}\bigg],
\\\nonumber
p_{r}&=&\frac{1}{6r^{6}\big(2b^{2}+b-1\big)(r-\nu)^{2}}
\bigg[\big(-10b
\mu ''r^{3}+8b\nu'r-10b\mu'^{2}\big(r^{2}-4 a_{1}\nu'\big)r
\\\nonumber
&+&\mu'
\big(4(b-3)r^{2}+5b\nu'\big(r^{2}+4a_{1}\nu'\big)\big)\big)
r^{5}+\nu\big
(40br^{3}a_{1}\mu'^{3}+6r^{2}\big(5br^{2}-4a_{1}
\\\nonumber
&\times&\big(5\nu'b+4b-2\big)\big)\mu'^{2}+r\big(-30b
a_{1}\nu'^{2}-2b\big(5r^{2}+62a_{1}\big)\nu'+r\big((36-17
\\\nonumber
&\times&b)r+20ba_{1}\big(4r\mu''+\nu''\big)\big)\big)\mu'-4(2b+3)a
_{1}\nu'^{2}+6r^{2}\big(5b\mu''r^{2}-2b+1\big)
\\\nonumber
&+&4br^{2}\nu'\big(5a_{1}\mu''-4\big)\big)r^{3}-\nu^{2}
\big(120br^{3}a_{1}\mu'^{3}+6r^{2}\big(5
br^{2}+a{1}\big(-20\nu'b-46
\\\nonumber
&\times&b+26\big)\big)\mu'^{2}+r\big(\big(-5\nu'b-22b
+36\big)r^{2}+2a_{1}\big(-5
b\nu'^{2}+(6-108b)\nu'
\\\nonumber
&+&4b\big(5r\big(6r\mu''+\nu''\big)-13\big)\big)\big)\mu'+(3-13b)
a_{1}\nu'^{2}+8\nu'\big(a_{1}\big(5br^{2} \mu''-6\big)
\\\nonumber
&-&br^{2}\big)+2r\big(6(1-2 b)r+\big(15
br^{2}+2(b+6)a_{1}\big)\mu''r+2(3-2b) a_{1}\nu''\big)\big)r^{2}
\\\nonumber
&+&\nu^{3}\big(\big(-12b+r\big(\mu'\big(10
r\mu'b-9b+12\big)+10br\mu''\big)+6\big)r^{2}+2a_{1}\big(60br^{3}\mu'^{3}
\\\nonumber
&-&4r^{2}\big(5\nu'b+33b-21\big)\mu'^{2}+r\big(10r\big(12 r\mu
''+\nu''\big)b-93b+(6-46b)\nu'
\\\nonumber
&+&6\big)\mu'+4b+\nu'\big(10b\mu''r^{2}-5 b-9\big)+2r\big(2(b+6)
r\mu''+(3-2b)\nu''\big)
\\\nonumber
&-&18\big)\big)r+a_{1}\nu^{4}\big(-4\big(20r\mu'b+b+6\big)
\mu''r^{2}-2\mu'\big(-41b+2
r\mu'\big(10r\mu'b
\\\nonumber
&-&21b+15\big)+6\big)r-3 (b-7)\big)\bigg],
\\\nonumber
p_{\bot}&=&\frac{1}{6\big(2b^{2}+b-1\big) r^{6}(r-\nu
)^{2}}\bigg[\big(2(b-3)\mu''r^{3}+(2 b+3)\nu'r+2(b-3)\mu'^{2}
\big(r^{2}
\\\nonumber
&-&4a_{1}\nu'\big)r-\mu'\big(2(4
b+3)r^{2}+(b-3)\nu'\big(r^{2}+4a_{1}\nu'\big)\big)\big)r^{5}-\nu
\big(8(b-3)r^{3}a_{1}
\\\nonumber
&\times&\mu'^{3}+6r^{2}\big((b-3)r^{2}-4
a_{1}\big(2b+(b-3)\nu'-1\big)\big)\mu'^{2}+r\big(-6(b-3)a_{1}\nu'^{2}
\\\nonumber
&+&\big(4(b+15)a_{1}-2
(b-3)r^{2}\big)\nu'+r\big(4(b-3)a_{1}\big(4r\mu
''+\nu''\big)-5(5b+3)
\\\nonumber
&\times&r\big)\big)\mu'+2(10 b+3)
a_{1}\nu'^{2}+3r^{2}\big(2(b-3)\mu''r^{2}-2b+1\big)+2r^{2}\nu'\big(2b+2
\\\nonumber
&\times&(b-3)a_{1}\mu''+3\big)\big)r^{3}+\nu^{2}\big(24(b-3)r^{3}a_{1}\mu'^{3}+6
r^{2}\big((b-3)r^{2}+2a_{1} \big(-13
\\\nonumber
&\times&b-2(b-3)\nu'+5\big)\big)\mu'^{2}+r\big(2
a_{1}\big(-(b-3)\nu'(r)^{2}+48\nu'+4\big(b+(b-3)
\\\nonumber
&\times&r\big(6 r\mu''+\nu''\big)+6\big)\big)-r^{2}\big(26
b+(b-3)\nu'+12\big)\big)\mu'+(b+3)a_{1}\nu'^{2}+\nu'
\\\nonumber
&\times&\big((2 b+3)r^{2}+4a_{1}\big(2 (b-3)\mu ''r^{2}+18
b+3\big)\big)+2r\big(r \big(-6 b+\big(3(b-3)
\\\nonumber
&\times&r^{2}-2(7b+3) a_{1}\big)\mu''+3\big)-8b a_{1}
\nu''\big)\big)r^{2}+\nu^{3}\big(r^{2}\big(6b+r\big(\mu' \big(9b-2(b
\\\nonumber
&-&3)r\mu'+3\big)-2(b-3) r\mu''\big)-3\big)-2a_{1}
\big(12(b-3)r^{3}\mu'^{3}-4r^{2}\big(21b+(b
\\\nonumber
&-&3)\nu'-6\big)\mu'^{2}+
r\big(3(b+13)-2(b-9)\nu'+2(b-3)r\big(12r\mu''+\nu''\big)\big)\mu'
\\\nonumber
&+&26b-4(7b+3)
r^{2}\mu''+\nu'\big(2(b-3)\mu''r^{2}+17b+3\big)-8br\nu''+3\big)\big)r
\\\nonumber
&+&a_{1}\nu^{4}
\big(33b+2r\big(\mu'\big(-b+2r\mu'\big(-15b+2(b-3)r\mu'+3\big)+15\big)+2r
\\\nonumber
&\times&\big(-7b+4(b-3 )r\mu'-3\big)\mu''\big)+3\big)\bigg].
\end{eqnarray}

\section*{Appendix B: Field Equations (Model 2)}

The resulting filed equations corresponding to model 2 are
\begin{eqnarray}\nonumber
\rho&=&\frac{1}{6\big(2b^{2}+b-1\big)r^{2}\nu^{3}
\big(\nu\big(2r\mu'+1\big)-r\big(2r\mu'+\nu
'\big)\big)^{3}}\bigg[a_{2}r^{5}(r-\nu)\big(-20br^{5}
\\\nonumber
&\times&\mu'\nu'\big(2r\mu'+\nu'\big)+r^{3}\nu
\big(20br^{3}\mu'^{3}+2\nu'\big(5
br^{2}\mu''+(3-2b)\nu'\big)+120br^{2}\mu'^{2}
\\\nonumber
&\times&\big(\nu'+1\big)+r\mu'\big(-20br\big(r\mu''+\nu''\big)
+15b\nu'^{2}+4(28
b+3)\nu'\big)\big)-r^{2}\nu^{2}\big(60
\\\nonumber
&\times&br^{3}\mu'^{3}+\nu'\big(20br^{2}\mu''
-36b+54\big)+12r^{2}\mu'^{2}\big(10b\nu'+31
b+1\big)+2r\big((13b
\\\nonumber
&-&12)r\mu''+2(2b-3)\nu''\big)+r\mu'\big(-20 br\big(3
r\mu''+2\nu''\big)-5b\nu'^{2}+6(31b+6)
\\\nonumber
&\times&\nu'+68b+48\big)+(7b-3)\nu'^{2}\big)+r\nu^{3}\big(60br^{3}
\mu'^{3}+2\big(2\big((13b-12)r^{2}\mu''
\\\nonumber
&+&(2b-3)\big(r\nu''-4\big)\big)+\nu'\big(5b
r^{2}\mu''-7b+18\big)\big)+8r^{2}\mu'^{2}\big(5b\nu'+48b+3\big)
\\\nonumber
&+&r\mu'\big(-20br
\big(3r\mu''+\nu''\big)+(74b+24)\nu'+3(41b+36)\big)\big)
+\nu^{4}\big(r\big
(2r\mu''
\\\nonumber
&\times&\big(10br\mu'-13b+12\big)-\mu'\big(4r\mu'\big(5br\mu'+33
b+3\big)+55b+60\big)\big)+21b
\\\nonumber
&-&39\big)\big)-\nu^{2}\big(\nu\big(2 r\mu'+1\big)-r
\big(2r\mu'+\nu'\big)\big)^{2}\big(3a_{2}(b+1)r^{5}
(r-\nu)+\nu\big(\nu
\big(2r
\\\nonumber
&\times&\mu'+1\big)-r\big(2r\mu'+\nu'\big)\big)
\big(10br(\nu-r)\mu''+5b\mu'\big(r\big(\nu'-4\big)
+3\nu
\big)+10br
\\\nonumber
&\times&(\nu-r)\mu'^{2}+(6-4b)\nu'\big)\big)\bigg],
\\\nonumber
p_{r}&=&-\frac{1}{6\big(2b^{2}+b-1\big)r^{3}\nu^{3}\big(\nu
\big(2r\mu'+1\big)-r\big(2
r\mu'+\nu'\big)\big)^{3}}\bigg[a_{2}r^{6}(r-\nu) \big(-20
\\\nonumber
&\times&br^{5}\mu'\nu'\big(2r\mu'+\nu'\big)+r^{3}\nu\big(20b
r^{3}\mu'^{3}+2\nu'\big(5br^{2}\mu''+(6-8 b)\nu'\big)+24r^{2}
\\\nonumber
&\times&\mu'^{2}\big(5b\nu'+3b+1\big)+r\mu'\big(-20
br\big(r\mu''+\nu''\big)+15 b\nu'^{2}+4(16b+9)\nu'\big)\big)
\\\nonumber
&-&r^{2}\nu^{2}\big(6
br^{3}\mu'^{3}+\nu'\big(20br^{2}\mu''-60b+66\big)
+12r^{2}\mu'^{2}\big(10b\nu'+19
b+7\big)
\\\nonumber
&+&2r\big((13b-12)r\mu''+2 (2b-3)\nu''\big)
+r\mu'\big(-20br\big(3
r\mu''+2\nu''\big)-5b\nu'^{2}
\\\nonumber
&+&(90b+84)\nu'+20b+72\big)+(3-5
b)\nu'^{2}\big)+r\nu^{3}\big(60br^{3}\mu'^{3}+2
\big(\nu'\big(5br^{2}\mu''
\\\nonumber
&-&19b+24\big)+2r\big((13b-12)r\mu''+(2
b-3)\nu''\big)-22b+27\big)+8r^{2}\mu'^{2}\big(5
\\\nonumber
&\times&b\nu'+30 b+12\big)+r\mu'
\big(-20br\big(3r\mu''+\nu''\big)+(26b+48)\nu'
+3(9b+52)\big)\big)
\\\nonumber
&+&\nu^{4}\big(r\big(\mu'\big(-4r\mu'\big(5br\mu'+21b+9\big)-7
(b+12)\big)+2r\mu''\big(10br\mu'-13b
\\\nonumber
&+&12\big)\big)+33 b-45\big)\big)-\nu^{2}\big(\nu\big(2
r\mu'+1\big)-r\big(2r\mu'+\nu'\big)\big)^{2}
\big(3a_{2} (b+1)r^{6}(r
\\\nonumber
&-&\nu)+\nu\big(\nu\big(2
r\mu'+1\big)-r\big(2r\mu'+\nu'\big)\big)\big(b
r\big(r\big(\mu'\big(-10r\mu'+5\nu'+4\big)-10r
\\\nonumber
&\times& \mu''\big)+8\nu'\big)+\nu\big(r\big(10br
\mu''+\mu'\big(10br\mu'-9b+12\big)\big)-12b+6\big)-12
\\\nonumber
&\times& r^{2}\mu'\big)\big)\bigg],
\\\nonumber
p_{\bot}&=&\frac{1}{6\big(2b^{2}+b-1\big)r^{3}\nu^{3}
\big(\nu\big(2r\mu'+1\big)-r\big(2r
\mu'+\nu'\big)\big)^{3}}\bigg[(a_{2}r^{6}(r-\nu)\big(-4
\\\nonumber
&\times&(b-3)r^{5}\mu' \nu'\big(2r\mu'+\nu'\big)+r^{3}\nu\big(4
(b-3)r^{3}\mu'^{3}+\nu'\big(2(b-3)r^{2}\mu''+(3
\\\nonumber
&-&14 b) \nu'\big)+12r^{2}\mu'^{2}\big(2(b-3)\nu'-5\big)-r\mu'
\big(4(b-3)r\big(r\mu''+\nu''\big)-3(b
\\\nonumber
&-&3)\nu'^{2}+4(4b+15)\nu'\big)\big)+r^{2}\nu^{2}\big(-12
(b-3)r^{3}\mu'^{3}+2\nu'\big(-2(b-3)r^{2}\mu''
\\\nonumber
&+&42b-3\big)+12r^{2}
\mu'^{2}\big(-2(b-3)\nu'+b+16\big)+2r\big((3-17b)
r\mu''-8b\nu''\big)
\\\nonumber
&+&r\mu'\big(4\big((b-3)r\big(3
r\mu''+2\nu''\big)+17b+12\big)+(b-3)\nu'^{2}+6(9b+17)\nu' \big)
\\\nonumber
&+&b\nu'^{2}\big)+r\nu^{3}\big(12(b-3)r^{3}\mu'^{3}
+2\nu'\big((b-3)r^{2}\mu''-29
b\big)+4r^{2}\mu'^{2}\big(2(b-3)
\\\nonumber
&\times&\nu'-6b-51\big)+4r\big((17b-3)
r\mu''+4b\nu''\big)-r\mu'\big(4(b-3)r\big(3r \mu''+\nu''\big)
\\\nonumber
&+&(38b+42)\nu'+153b+93\big)-70
b+3\big)+\nu^{4}\big(r\big(\mu'\big(4r\mu'\big(3 (b+6)-(b
\\\nonumber
&-&3)r\mu'\big)+85 b+45\big)+2r\mu''\big(2
(b-3)r\mu'-17b+3\big)\big)+57b\big)\big)+\nu^{2}\big(\nu \big(2
\\\nonumber
&\times&r\mu'+1\big)-r\big(2r
\mu'+\nu'\big)\big)^{2}\big(3a_{2}(b+\mu'+1\big)
-r\big(2r\mu'+\nu'\big)\big)\big(r
\big(-2(b
\\\nonumber
&-&3)r^{2}\mu''-2(b-3)r^{2}
\mu'^{2}+r\mu'\big((b-3)\nu'+8b+6\big)-(2 b+3)
\nu'\big)+\nu\big(r
\\\nonumber
&\times&\big(2(b-3)r\mu''+\mu'\big(2(b-3)r\mu'-9
b-3\big)\big)-6b+3\big)\big)\big)\bigg].
\end{eqnarray}

\section*{Appendix C: Field Equations (Model 3)}

The corresponding $f(\textmd{Q},T)$ field equations are
\begin{eqnarray}\nonumber
\rho&=&\frac{1}{6(b+1)(2b-1)r^{3}(r-\nu)}\bigg[\big(10bJ
\mu'^{2}r^{2}+\big(10b J \mu''-3(b+1)\big(k -\frac{1}{r^{2}}
\\\nonumber
&\times&(\nu\big(2\mu'{r^{2}}\frac{r\nu'-\nu}{r(r-\nu
)}\big)\big)\big)\big)r^{2}+5 b\big(\frac{1}{\nu^{4}(L)^{4}}\big(2
\text{k r}^{9}a_{3}\big(a_{3}-\frac{2\nu L}{r^{2}}\big)
{\nu^{4}L\big)^{4}}\big(\frac{2L \nu}{r^{3}}
\\\nonumber
&-&\frac{1}{r^{2}}\big(\big(-\frac{r\nu'-\nu}{r^{2}
(r-\nu)}-\frac{\big(1-\nu'\big)\big(r\nu'-\nu\big)}{r (r-\nu)^{2}}
+2\mu''+\frac{\nu''}{r-\nu
}\big)\nu\big)-\frac{\nu'L}{r^{2}}\big)\big)
\\\nonumber
&\times& -\big(\nu'-4\big)J\big) \mu'r+2(2 b-3)J\nu'\big) r^{2}
+\big(-\frac{1}{\nu^{4}\big(L\big)^{4}}\big(2ka_{3}\big(10 r\mu'b-2b
\\\nonumber
&\times&+3\big)\big(a_{3}-\frac{2\nu L}{r^{2}}\big)
\big(\frac{2L\nu}{r^{3}}-\frac{1}{r^{2}}
\big(\big(-\frac{r\nu'-\nu}{r^{2}
(r-\nu)}-\frac{\big(1-\nu'\big) \big(r
\nu'-\nu\big)}{r(r-\nu)^{2}}+2\mu''
\\\nonumber
&+&\frac{\nu''}{r-\nu
}\big)\nu\big)-\frac{\nu'L}{r^{2}}\big)r^{9}\big)+3(b+1)
\big(k-\frac{\nu L}{r^{2}}\big)r^{2}-J\big(20b\mu'^{2}r^{2}+20b
\mu''r^{2}
\\\nonumber
&+&\mu'\big(-5\nu'b+41b+6\big)r+(7b-3)\nu'\big)\big)
\nu r+\big(3
(b+1)J+r\big(10bJ\mu''r
\\\nonumber
&\times&+J\mu'\big(10r\mu'b
+21b+6\big)+\frac{1}{\nu^{4}\big(L\big)^{4}}\big(2kr^{8}a_{3}
\big(5r\mu'b-2b+3\big)\big(a_{3}-\frac{2\nu
L}{r^{2}}\big)
\\\nonumber
&\times&
\big(\frac{2L\nu}{r^{3}}-\frac{1}{r^{2}}
\big(\big(-\frac{r\nu'-\nu
}{r^{2}(r-\nu)}
-\frac{\big(1-\nu'\big)\big(r\nu'-\nu\big)}{r(r-\nu)^{2}}+2
\mu''+\frac{\nu''}{r-\nu}\big)\nu\big)
\\\nonumber
&-&\frac{\nu'L}{r^{2}}\big)\big)\big)\big)\nu^{2}\bigg],
\\\nonumber
p_{r}&=&\frac{1}{6(b+1)(2b-1)r^{3}(r-\nu)}\big(r\nu
\big(\frac{1}{L^{4}\nu^{4}}\big(2a_{3}\text{K
r}^{9}\big(a_{3}-\frac{2L\nu}{r^{2}}\big) \big(10br\mu'-2b
\\\nonumber
&+&3\big)\big(\frac{2L\nu}{r^{3}}-\frac{L\nu'}{r^{2}}
-\frac{1}{r^{2}}\big(\nu\big(-\frac{r\nu'-\nu}{r^{2}(r-\nu)}+2
\mu''+\frac{\nu''}{r-\nu}-\frac{1}{r (r-\nu)^{2}}\big(\big(1
\\\nonumber
&-&\nu'\big)\big(r\nu'-\nu\big)\big)\big)\big)\big)
\big)+J\big(20br^{2}
\mu''+20br^{2}\mu'^{2}+r\mu'\big(-5b\nu'-7 b+30\big)
\\\nonumber
&+&(3-5b)\nu'\big)+(6-12b)J-3(b+1)r^{2}
\big(K-\frac{L\nu}{r^{2}}\big)\big)+r^{2}\big(r\mu'\big(5b
\big(J\nu'
\\\nonumber
&-&\frac{1}{L^{4}\nu^{4}}\big(2 a_{3}Kr^{9} \big(a_{3}
-\frac{2L
\nu}{r^{2}}\big)\big(\frac{2L\nu}{r^{3}}-\frac{L
\nu'}{r^{2}}-\frac{1}{r^{2}}\big(\nu
\big(-\frac{r\nu'-\nu}{r^{2}(r-\nu)}+2\mu''
\\\nonumber
&+&\frac{\nu''}{r-\nu}-\frac{\big(1-\nu'\big)
\big(r\nu'-\nu\big)}{r
(r-\nu)^{2}}\big)\big)\big)\big)\big)+4(b-3)J\big)
+r^{2}\big(3(b+1)\big(K
\\\nonumber
&-&\frac{L\nu}{r^{2}}\big)-10b J\mu''\big)-10bJr^{2}\mu'^{2}+8bJ
\nu'\big)+\big(-\nu^{2}\big)\big(\frac{1}{L^{4}\nu^{4}}
\big(2a_{3}Kr^{9}\big(a_{3}
\\\nonumber
&-&\frac{2 L\nu}{r^{2}}\big)\big(5br\mu'-2 b+3\big)
\big(\frac{2L\nu}{r^{3}}-\frac{L\nu'}{r^{2}}-\frac{1}{r^{2}}\big(\nu
\big(-\frac{r\nu'-\nu}{r^{2}(r-\nu)}+2\mu''
\\\nonumber
&+&\frac{\nu''}{r-\nu}-\frac{\big(1-\nu'\big)\big(r\nu'-\nu\big)}{r
(r-\nu)^{2}}\big)\big)\big)\big)+J\big(r\big(10br \mu''+\mu '(r)
\big(10br\mu'-3 b
\\\nonumber
&+&18\big)\big)-9 b+9\big)\big)\bigg],
\\\nonumber
p_{\bot}&=&\frac{1}{6(b+1)(2b-1)r^{3}(r-\nu)}
\bigg[r^{2}\big(r
\mu'\big(\frac{1}{L^{4}\nu^{4}}\big(2a_{3}(b-3)Kr^{9}
\big(a_{3}-\frac{2L
\nu}{r^{2}}\big)
\\\nonumber
&\times&\big(\frac{2L\nu}{r^{3}}-\frac{L\nu'}{r^{2}}
-\frac{\nu\big(-\frac{r\nu'-\nu}{r^{2}
(r-\nu)}+2\mu''+\frac{\nu''}{r-\nu}-\frac{\big(1-\nu
'\big)\big(r\nu'-\nu\big)}{r(r-\nu)^{2}}\big)}{r^{2}}
\big)\big)-J
\big((b
\\\nonumber
&-&3)\nu'+8b+6\big)\big)+r^{2}\big(2(b-3)
J\mu''+3(b+1)\big(K-\frac{L\nu}{r^{2}}\big)\big)+2(b-3)
\\\nonumber
&\times&Jr^{2}\mu'^{2}+(2b+3)J\nu'\big)+r\nu\big(\frac{1}
{L^{4}\nu^{4}}\big(4a_{3}Kr^{9}\big(a_{3}-\frac{2
L\nu}{r^{2}}\big)\big(2b-(b-3)r
\\\nonumber
&\times&\mu'\big)\big(\frac{2L
\nu}{r^{3}}-\frac{L\nu'}{r^{2}}-\frac{1}{r^{2}}\big(\nu
\big(-\frac{r\nu'-\nu}{r^{2}(r-\nu)}+2
\mu''+\frac{\nu''}{r-\nu}-\frac{1}{r (r-\nu)^{2}}\big(\big(1
\\\nonumber
&-&\nu'\big)\big(r\nu'-\nu\big)\big)\big)\big)\big)\big)
+J\big(-4(b-3)r^{2}\mu''-4
(b-3)r^{2}\mu'^{2}+r\mu'\big((b-3)
\\\nonumber
&\times&\nu'+23b+15\big)+b\nu'\big)+(6b-3)J-3
(b+1)r^{2}\big(K-\frac{L\nu}{r^{2}}\big)\big)+\nu^{2}\big(r
\big(\frac{1}{L^{4}\nu^{4}}
\\\nonumber
&\times&\big(2a_{3}
Kr^{8}\big(a_{3}-\frac{2L\nu}{r^{2}}\big)\big((b-3)r\mu'-4b\big)
\big(\frac{2L\nu}{r^{3}}-\frac{L\nu'}{r^{2}}
-\frac{1}{r^{2}}\big(\nu\big(-\frac{r\nu'-\nu}{r^{2}(r-\nu)}
\\\nonumber
&+&2\mu''+\frac{\nu ''}{r-\nu}-\frac{\big(1-\nu'\big)
\big(r\nu'-\nu\big)}{r(r-\nu)^{2}}\big)\big)\big)\big)+2(b-3)Jr
\mu''+J\mu'\big(2(b-3)r\mu'
\\\nonumber
&-&-3(5b+3)\big)\big)-9bJ\big)\bigg],
\end{eqnarray}
where
\begin{eqnarray}\nonumber
J&=&1-\frac{a_{3} r^{4} \exp \big(-\frac{a_{3} r^{2}}{\nu\big(2
\mu'+\frac{r\nu'-\nu}{r (r-\nu)}\big)}\big)}{\nu^{2}
\big(2\mu'+\frac{r\nu'-\nu}{r (r-\nu)}\big)^2}, \quad k
=\exp\big(-\frac{a_{3}r^{2}}{\nu\big(2\mu'+\frac{r \nu '-\nu}{r
(r-\nu)}\big)}\big),
\\\nonumber
L&=&\big(2\mu'+\frac{r\nu'-\nu}{r(r-\nu)}\big).
\end{eqnarray}
\\
\textbf{Data Availability:} No data was used for the research
described in this paper.

\end{document}